\documentclass[a4paper,11pt]{article}
\usepackage{subcaption}

\pdfoutput=1 

\usepackage{jinstpub} 

\usepackage{xspace}
\usepackage{siunitx}
\usepackage{multirow}
\usepackage{lineno}
\newcolumntype{C}[1]{>{\centering\let\newline\\\arraybackslash\hspace{0pt}}m{#1}}

\title{Design and Calibration of an Optically Segmented Single 
Volume Scatter Camera for Neutron Imaging}

\author[a,1]{A.~Galindo-Tellez\begin{NoHyper}\note{Corresponding author.}\end{NoHyper},}
\author[a]{K.~Keefe,}
\author[a]{E.~Adamek,}
\author[b]{E.~Brubaker,}
\author[a]{B.~Crow,}
\author[a]{R.~Dorrill,}
\author[a]{A.~Druetzler,}
\author[a]{C.J.~Felix,}
\author[a]{N.~Kaneshige,}
\author[a]{J.G.~Learned,}
\author[c]{J.J.~Manfredi,}
\author[a]{K.~Nishimura,}
\author[a]{B.~Pinto Souza,}
\author[a]{D.~Schoen,}
\author[b]{and M.~Sweany}

\affiliation[a]{Department of Physics and Astronomy, University of Hawai`i at M\={a}noa, \\Honolulu, HI 96822, USA}
\affiliation[b]{Sandia National Laboratories, \\Livermore, CA 94550, USA}
\affiliation[c]{Department of Nuclear Engineering, University of California, Berkeley, \\Berkeley, CA 94720, USA}

\emailAdd{alinegt@hawaii.edu}

\abstract{The Optically Segmented Single Volume Scatter Camera (OS-SVSC) aims to image neutron sources for non-proliferation applications using the kinematic reconstruction of elastic double-scatter events. Our prototype system consists of 64 EJ-204 organic plastic scintillator bars, each measuring $\SI{5}{mm}$ $\times$ $\SI{5}{mm}$ $\times$ $\SI{200}{mm}$ and individually wrapped in Teflon tape. The scintillator array is optically coupled to two silicon photomultiplier ArrayJ-60035 64P-PCB arrays, each comprised of 64 individual $\SI{6x6}{mm}$ J-Series sensors arranged in an 8 $\times$ 8 array.

We report on the design details, including component selections, mechanical design and assembly, and the electronics system. The described design leveraged existing off-the-shelf solutions to support the rapid development of a phase 1 prototype. Several valuable lessons were learned from component and system testing, including those related to the detector's  mechanical structure and electrical crosstalk that we conclude originates in the commercial photodetector arrays and the associated custom breakout cards.

We detail our calibration efforts, beginning with calibrations for the electronics, based on the IRS3D application-specific integrated circuits, and their associated timing resolutions, ranging from \SIrange{30}{90}{ps}. With electronics calibrations applied, energy and position calibrations were performed for a set of edge bars using $^{22}$Na and $^{90}$Sr, respectively, reporting an average resolution of \SI{12.07\pm0.03}{mm} for energy depositions between \SI{900}{keVee} and \SI{1000}{keVee}. We further demonstrate a position calibration method for the internal bars of the matrix using cosmic-ray muons as an alternative to emission sources that cannot easily access these bars, with an average measured resolution of \SI{14.86+-0.29}{mm} for depositions between \SI{900}{keVee} and \SI{1000}{keVee}. The coincident time resolution reported between pairs of bars measured up to $\SI{400}{ps}$ from muon acquisitions. Energy and position calibration values measured with muons are consistent with those obtained using particle emission sources.
}

\keywords{Neutron detectors (fast neutrons), 
Search for radioactive and fissile materials, 
Detector alignment and calibration methods (lasers, sources, particle-beams)}

\begin{document}
\maketitle
\flushbottom

\section{Introduction}
Neutron scatter cameras use the kinematic information from two elastic scatter interactions in hydrogenous material to reconstruct the neutron source direction and energy. This technique has been proposed for non-proliferation applications to image Special Nuclear Materials with neutron energies in the $\sim$\,\si{\mega\electronvolt} range (e.g.,~\cite{bib:Mascarenhas,bib:Marleau,bib:Riviere}), as well as in astronomical applications to image neutrons of a few to hundreds of \si{\mega\electronvolt}~\cite{bib:Grannan,bib:Wiinderer,bib:Zych}.

While traditional neutron scatter cameras have used a dual plane design
that limits acceptance, omnidirectional imaging has been obtained in a 
portable form factor with a height offset cylindrical 
configuration~\cite{bib:Goldsmith}. Despite such improvements, double scatter
imaging generally suffers from poor efficiency due to the low probability 
of multiple neutron scatters occurring in spatially separated scintillation cells.

Multiple concepts have been proposed to enhance the efficiency over previous
designs, with a focus on moving toward a single volume
scintillation region, rather than on discrete, spatially distant 
scintillators. A compact active volume dramatically improves the likelihood
that a neutron will scatter twice within the active scintillation volume.

Truly monolithic designs propose to reconstruct multiple interaction 
positions within a single volume~\cite{bib:Braverman,bib:Jocher}. A second 
technique proposes closely packed but optically segmented
scintillators~\cite{bib:Weinfurther}. In both cases, fine spatial resolution
of \SI{\sim 10}{mm} and fine temporal resolution of \SI{<1}{ns} could
increase the efficiency by an order of magnitude, relative to existing neutron
scatter camera systems~\cite{bib:Braverman,bib:Weinfurther}. Both design
approaches are being actively studied within the Single Volume Scatter Camera
(SVSC) Collaboration, led by Sandia National Laboratories.

This work describes an Optically Segmented SVSC (OS-SVSC) design
of 64~Teflon-wrapped EJ-204 plastic
scintillator bars. Details of the active volume, the mechanical design, and the
electronics are described, with an emphasis on limitations and issues that may
arise in other similar detector concepts. We report calibration
measurements for a subset of bars from this device, including an alternative method for 
calibrating bars that are not easily accessible by an external source.

\subsection{Neutron Double-Scatter Imaging Methodology}
The imaging methodology of the OS-SVSC requires neutrons to scatter at least twice in different scintillator bars 
to allow an estimate of the incident neutron direction and energy~\cite{bib:Weinfurther}. In the first interaction, a neutron 
scatters off a proton and deposits part of its energy in the bar, which can be obtained by measuring the intensity 
of the light detected. Our collaboration has performed proton light yield measurements \cite{bib:Laplace}
which are used to convert the measured light into proton recoil energies. The outgoing energy of the second 
neutron scatter, $E_{n'}$, is calculated using the distance $d$ and the time of flight $\Delta t$ between the two 
scatters:
\begin{equation}
  E_{n'}=\frac{1}{2}m_{n}v^{2}=\frac{1}{2}m_{n}{\left(\frac{d}{\Delta t}\right)}^{2}\,,
\end{equation}
where $m_{n}$ is the mass of a neutron and $v$ is the speed of the scattered
neutron. This non-relativistic expression is valid for MeV-scale neutrons.
The initial incident
energy of the neutron $E_{n}$ can be obtained by adding up the outgoing energy
with the energy lost due to proton recoil $E_{p}$ in the first scatter as:
\begin{equation}
E_{n}=E_{n'}+E_{p}\,.
\end{equation}
The incident neutron direction is confined on a cone centered on the vector connecting the two scatters with opening angle $\theta$, as:
\begin{equation}
    \cos{\theta}=\sqrt{\frac{E_{n'}}{E_{n}}}\,.
\end{equation}

\begin{figure}[!h]
\centering
\includegraphics[width=.90\textwidth]{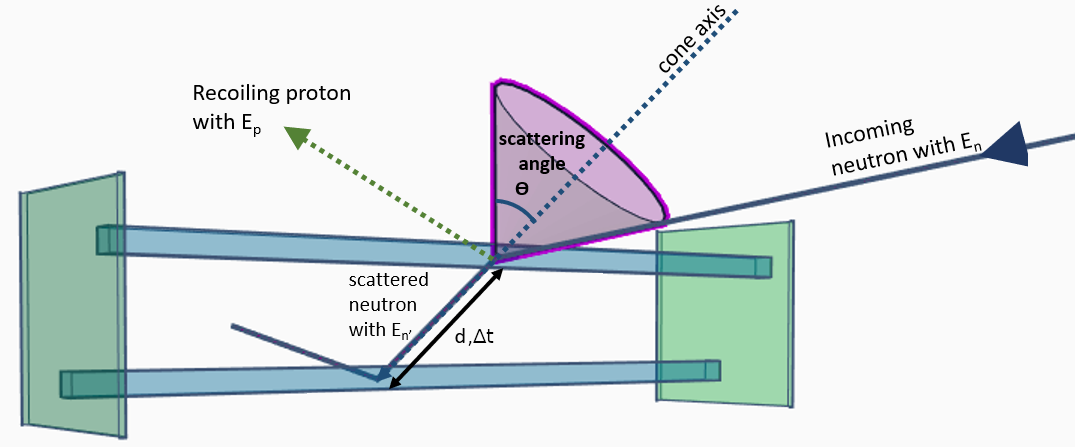}
\caption{Neutron double-scatter event illustration in the OS-SVSC. The projected
cone of one event is obtained using the time-of-flight of the neutron between
two bars and the intensity of the light in the bar of the first scatter.}
\label{doubleScatter}
\end{figure}

The physical process is illustrated in Fig.~\ref{doubleScatter}. An image of the neutron source can be obtained by overlapping back-projected cones to a plane placed at the estimated neutron source distance, or by implementing more
sophisticated iterative image reconstruction methods.

\section{OS-SVSC Prototype Design}
This section describes the design of the OS-SVSC, including scintillators and their surface wrapping, photodetectors, electronics and readout, and mechanical design. A primary motivation for many of the design choices was to use either existing in-house or readily available commercial parts in order to accelerate design and testing efforts. While this did lead to the first realization of a prototype, there are some detrimental impacts of these decisions, which are explained in detail in this section.

\subsection{Scintillator and Surface Wrapping}
The active volume of the OS-SVSC is composed of an array of 64 rectangular \SI{5}{mm} $\times$ \SI{5}{mm} $\times$ \SI{200}{mm} EJ-204 scintillator bars from Eljen~\cite{bib:Eljen}, each individually wrapped with three layers of Teflon tape. The decision to use this scintillator and surface treatment was motivated by our collaboration's previous measurements of interaction position and timing resolution using a variety of combinations of scintillating materials and surface treatments. A number of scintillators and surface wrappings were shown to give combined resolutions of better than \SI{10}{mm}, but the enhanced light output of EJ-204, combined with the ease of consistently wrapping Teflon relative to ESR, led to our particular chosen combination ~\cite{bib:Sweany}.

From experience with our single bar measurements across two different sites and experimental setups, there 
can be significant variation in the optical properties of Teflon due to variations in wrapping technique, the 
number of layers, and the underlying thickness and quality of the commercial Teflon tape. The standardization of 
the wrapping procedure, number of layers, and adoption of a minimum specification for procured Teflon (MIL 
SPEC T-27730A), all led to improved reproducibility of results and were adopted for the design of this detector. Three layers were necessary to achieve the desired thickness of approximately \SI{0.5}{mm}, which has been shown to maximize reflectivity~\cite{bib:Janecek}.

\subsection{Photodetectors, Optical Coupling, and Signal Conditioning}
The scintillator bars are optically coupled on each end to pixels of the ArrayJ-60035 64P-PCB arrays (ArrayJs), using individually cut squares of $\SI{5}{mm}\times\SI{5}{mm}\times\SI{0.5}{mm}$ Eljen EJ-560 silicone rubber optical interface~\cite{bib:Eljen2}. The reduced thickness of the optical interface relative to standard \SI{1}{mm} and \SI{2}{mm} offerings serves to improve the overall light yield for scintillation events. The use of individually cut pieces rather than a large single sheet of EJ-560 covering the entire array was selected to reduce the potential for optical crosstalk detected.

The use of the $8\times8$ array of \SI{6x6}{mm} SiPMs allows for a small cross-section, which is consistent with the compact OS-SVSC intended design. The underlying SiPM in the array is the same as that used in the single bar characterizations. Unlike our single bar measurements, which used the SensL J-Series SMA evaluation board to access the fast output (FOUT) of the SiPM, the SensL array requires some further circuitry to condition the FOUT for readout. A pair of commercial adapter cards (Ultralytics ArrayX-64 A.2) provide the same style readout as the SensL SMA evaluation board with identical schematic circuit but different component values without having a significant impact on the waveform. The readout is repeated for each of the 64~FOUTs in the ArrayJ. All FOUTs, along with a common bias line, are provided on a Samtec connector (QTE-040-03-F-D-A) that interfaces to the rest of the system via a Samtec EQCD cable (one per array).

The bias voltage is provided through a Keysight benchtop power supply, using an SMA
interface on the readout electronics.  It is carried back up through the 
EQCD cables to the ArrayJs.
Unless otherwise specified, the bias voltage for all operating conditions is 
set to \SI{30}{V}, in order to give the best possible timing 
performance \cite{bib:Dolinsky}.
Due to the relatively controlled temperature environment of the laboratory in which the OS-SVSC tests are 
carried out, we expect minor variations in gain of order 1\% to 5\%, using the same calculation as that found in 
~\cite{bib:Sweany2}. These variations are generally comparable to or smaller than those expected among 
individual SiPMs due to manufacturing variability~\cite{bib:SensL}.

\subsection{Mechanical Design}
\label{sect:mechanic}
\begin{figure}[!h]
\centering
\includegraphics[width=\textwidth]{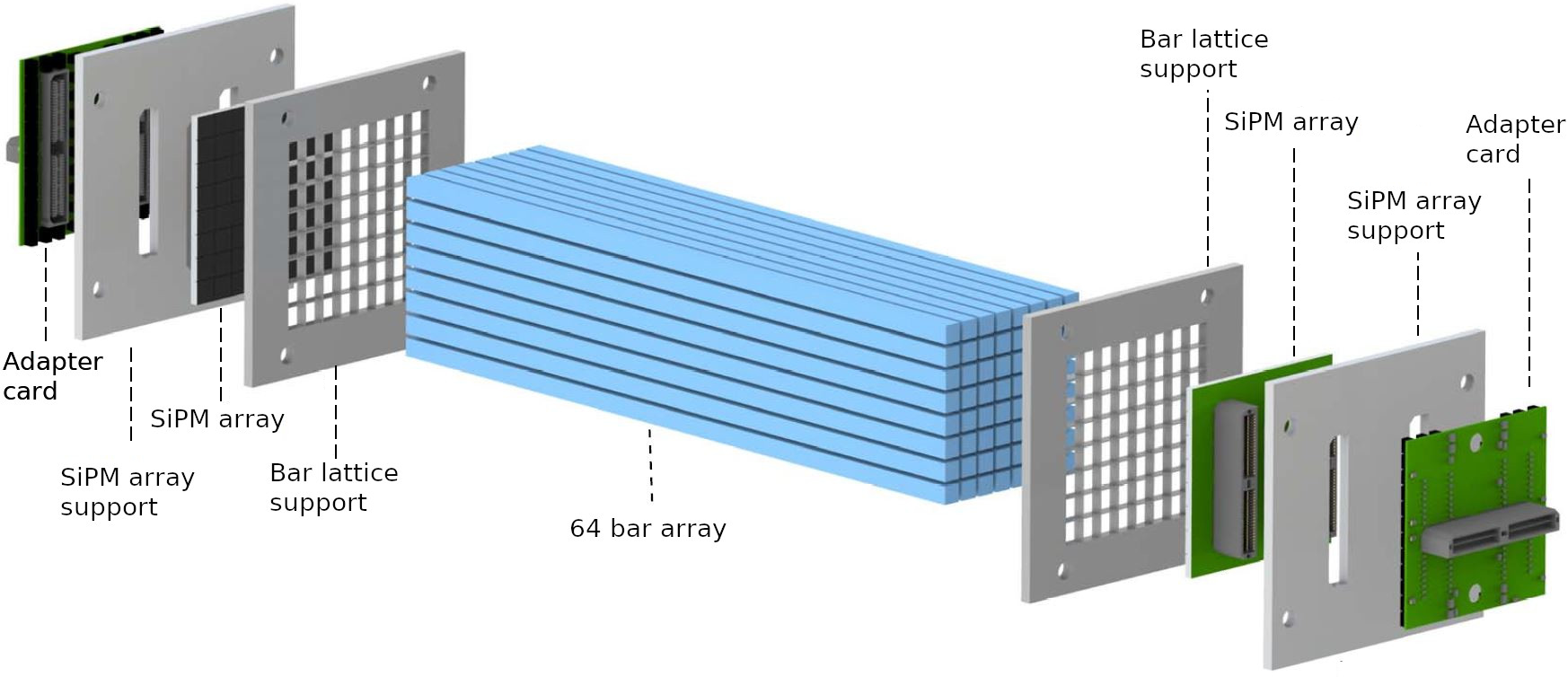}
\caption{Detector internals consisting of 64 bars (8$\times$8 array), two bar lattice
supports, two SensL SiPM arrays and two custom made adapter cards.}
\label{fig:mechanics3D}
\end{figure}

Design features, such as optical coupling, electrical 
connections/cabling, enclosure configuration, and internal
support details, were designed to match the ArrayJ dimensions and
layout. An illustration of the OS-SVSC detector internals is shown in
Fig.~\ref{fig:mechanics3D}.

The internal component supports and the aluminum enclosure were designed and fabricated in-house. The design 
of such components was motivated by the need to maintain alignment of individual pixels with the bars and to 
provide consistent and moderate contact pressure between the bars, the optical interface material, and each pixel's glass cover. With this, we assure maximum light transmission while reducing mechanical stress on the ArrayJ, which could lead to cracking of the glass. 
Two lattice supports were 3D printed with polylactic acid filament to align the Teflon wrapped bars with the 
SiPMs. In addition, two SensL SiPM array supports were 3D printed to maintain the ArrayJs in placement. The 
support designs are shown in Fig.~\ref{fig:mechanics3D}. All four supports are kept aligned using four threaded 
rods, which run the length of the detector, and spacers (tubes around the threaded rods) between all the 
supports.

\begin{figure}[!h]
\begin{subfigure}{.50\linewidth}
\includegraphics[width=\linewidth]{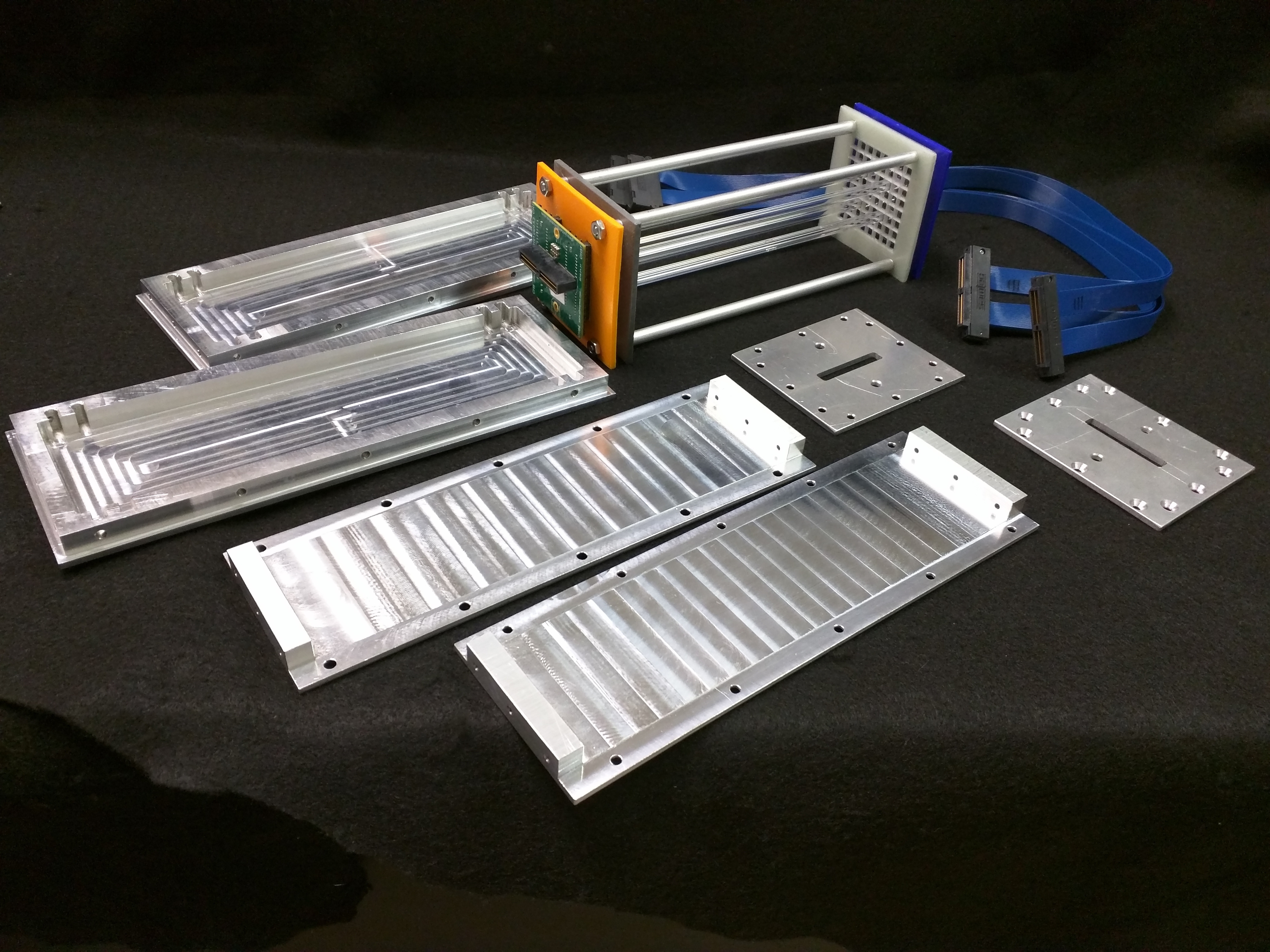}
\caption{}
\label{fig:mechanics_sub1}
\end{subfigure}%
\begin{subfigure}{.50\linewidth}
\centering
\includegraphics[width=\linewidth]{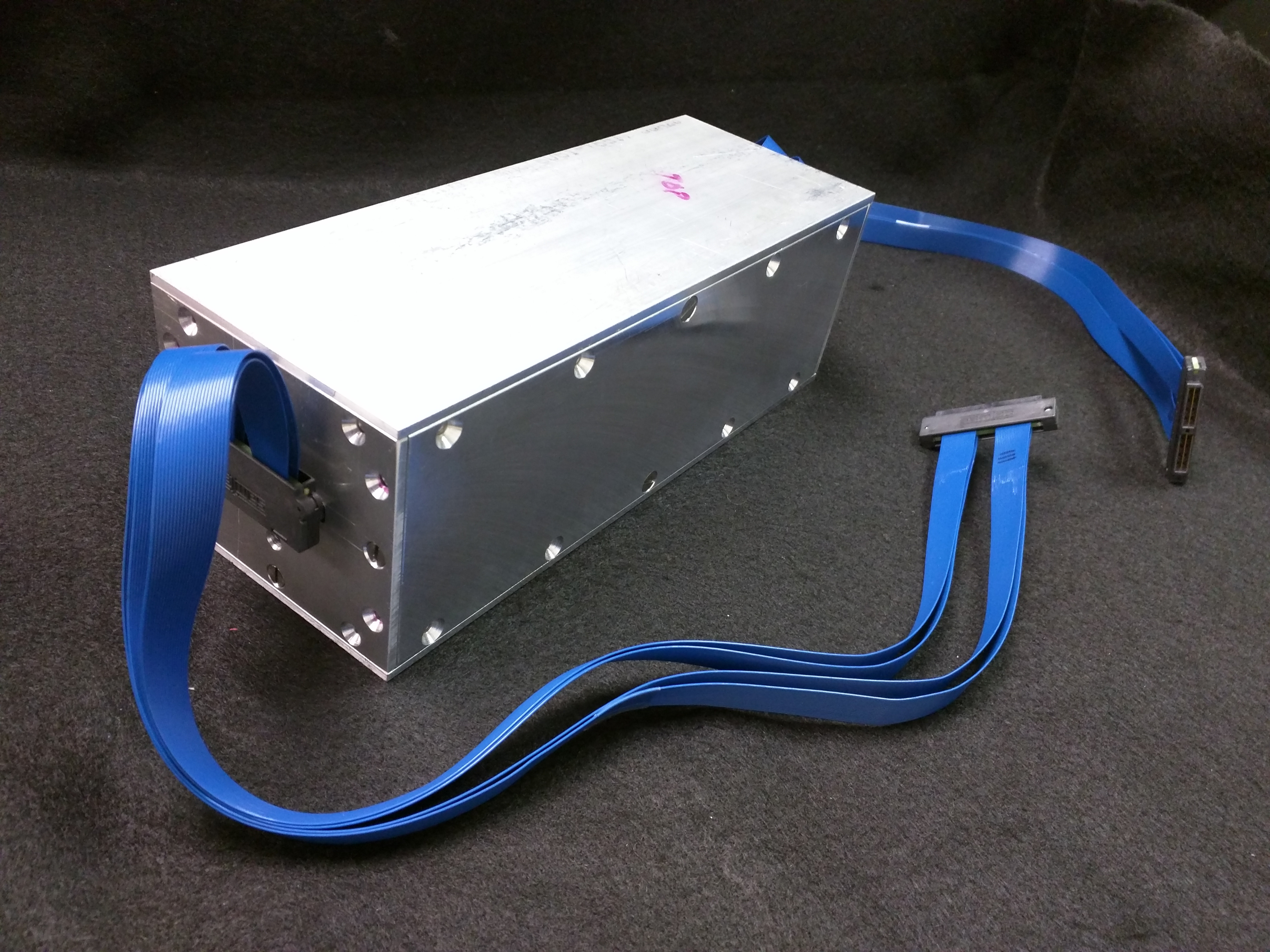}
\caption{Assembled OS-SVSC prototype enclosure.}
\label{fig:mechanics_sub2}
\end{subfigure}
\caption{Mechanical enclosure design for OS-SVSC prototype: (a) disassembled and (b) assembled.}
\label{fig:mechanics}
\end{figure}
Aluminum was chosen for the enclosure material for its low cost, and
compatibility with computer numerical control machining, as well as to
reduce the potential for neutron scatter in the enclosure.
The top and bottom sections have slots that match the supports to hold them in place. 
Rabbet joints and blind screw holes are used to minimize external light penetration. To avoid overheating issues 
and the need for active cooling, the electronics and power supply are connected to the ArrayJs with Samtec EQCD 
cables and reside outside of the enclosure. The assembled mechanical components without scintillator bars and 
ArrayJs are shown in Fig.~\ref{fig:mechanics}. Calibrations are performed with the detector in the horizontal orientation shown.
%

\subsection{Electronics and Data Acquisition}
The OS-SVSC design places constraints on the performance requirements of the electronics and data acquisition (DAQ) systems. It requires that the $8\times8\times2$ channels each be independently monitored for triggers and digitized with a time scale that can take advantage of the $\mathcal{O}$(100) ps coincident timing typical of the SiPMs \cite{bib:Sweany}. The digitization of each ArrayJ is carried out by independent standard control readout (SCROD) board stacks. For synchronizing the two SCRODs, a third external board called CAJIPCI, (Clock and JTAG\footnote{named after the Joint Test Action Group which codified it.} interface by peripheral component interconnect bus) is utilized to distribute a global clock. The SCRODs and CAJIPCI boards were previously utilized in the electronics system of the mini-TimeCube experiment and are described in more detail elsewhere~\cite{bib:Li}.

\begin{figure}[!h]
\centering
\includegraphics[width=0.7\textwidth]{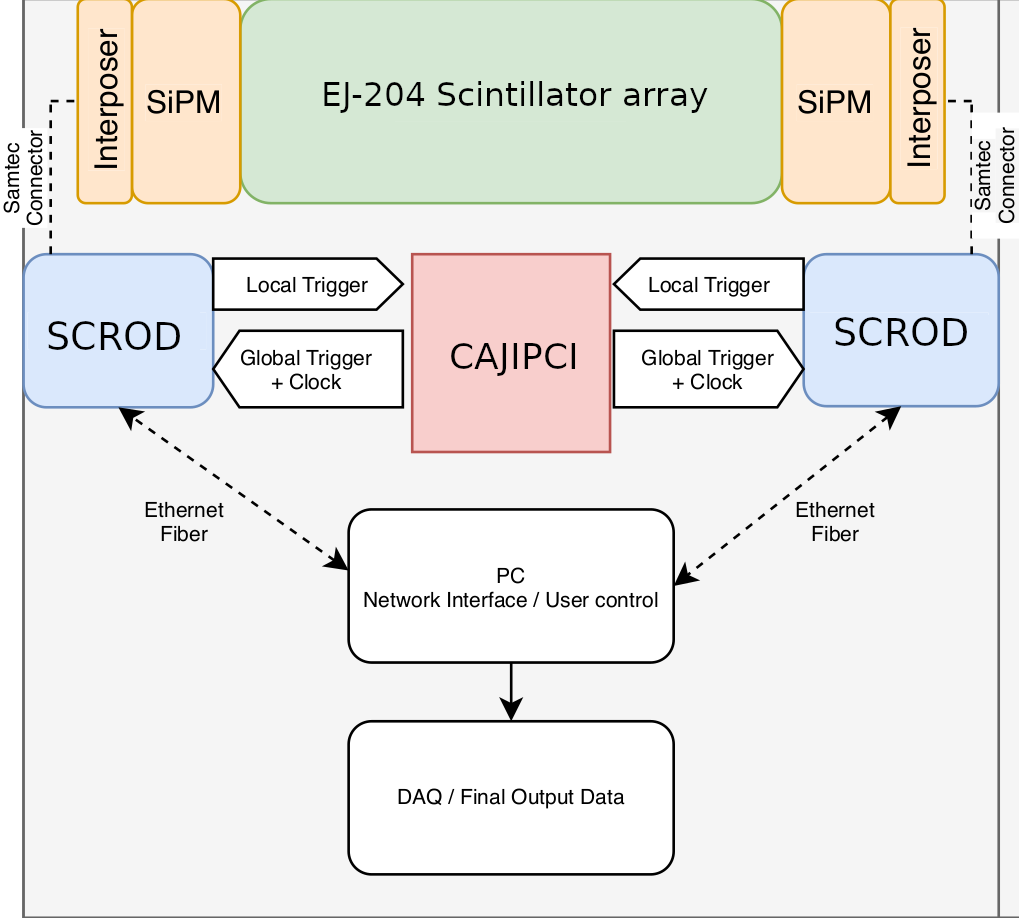}
\caption{Diagram of the front-end electronics and DAQ chain. The SCRODs connect via Samtec connector to the interposer cards from Ultralytics and are synchronized by the CAJIPCI board, responsible for the global trigger and clock. The user configures and reads data from the SCRODs via two parallel ethernet fibers.}
\label{front_end_electronics}
\end{figure}

\subsubsection{Front-End Electronics}
The front-end electronics functionality lies in the ice radio sampler (IRS), a family of application-specific integrated circuits (ASICs) developed for projects that require fast sampling and deep buffering~\cite{bib:Li}. A diagram of the electronics is shown in Fig.~\ref{front_end_electronics}. Features added for this system include improved networking interfaces control (containing MicroBlaze based networking), high-speed deserialization on trigger monitoring, adapter cards for carrier boards, Samtec connectors, and interposer cards for recording individual channels from the ArrayJ.

\subsubsection{Data Acquisition}
The user control between a PC and the SCRODs utilizes a User Datagram Protocol (UDP) interface to an embedded software system. Each SCROD, responsible for digitizing 64 channels on one edge of the detector, uses three UDP ports. Each port can be configured as input or output on an assigned IP address. The first output port is dedicated to the FPGA configuration, ASIC configuration, and digital-to-analog converter voltage settings. A second input port is used for digitizing each channel in an event. The third input port reads out trigger information, that indicates which channels we triggered during the event. Since UDP packet transmission is unreliable, the collected data must be immediately saved to prevent buffer overflow errors and eliminate packet loss. Each packet is sent with a unique message identification number in order to ensure each one can be uniquely identified and that events are complete with no missing packets.

After a completed run, the trigger information and digitized data are stored in independent binary files which are then combined into a single ROOT file~\cite{bib:ROOT} for further analysis. Electronics calibrations explained in Sec.~\ref{sec:elec_calib} are applied during the conversion from binary to ROOT data formats. In addition, metadata are stored separately for each board stack's register configuration. 

\section{Electrical Crosstalk}
\label{sec:crosstalk}
\begin{figure}[!h]
\centering
\begin{subfigure}{.45\textwidth} \centering 
\includegraphics[width=\linewidth]{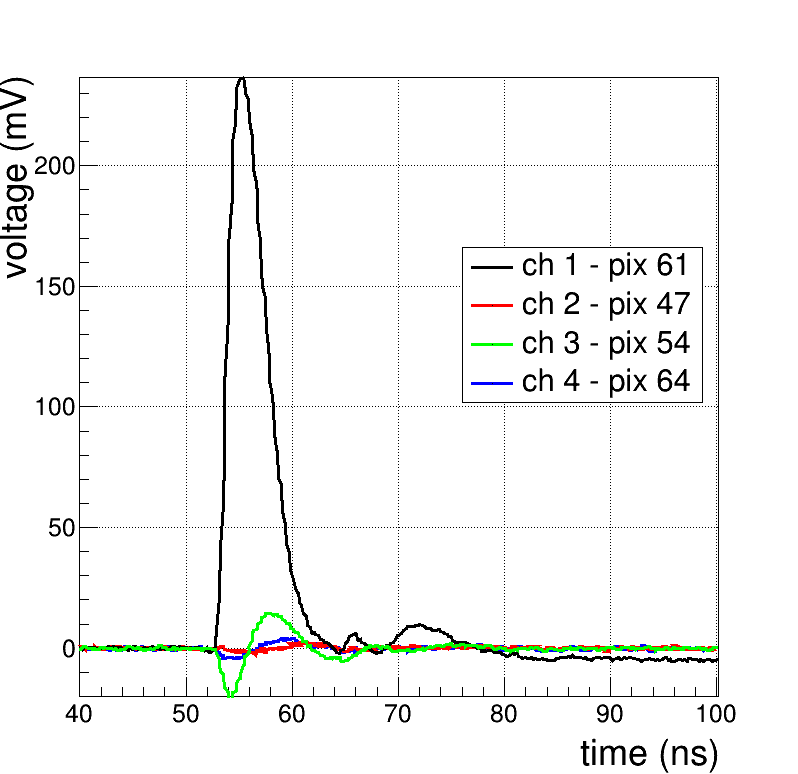}
\caption{}
\label{fig:crosstalk-fast}
\end{subfigure}
\begin{subfigure}{.45\textwidth} \centering 
\includegraphics[width=\linewidth]{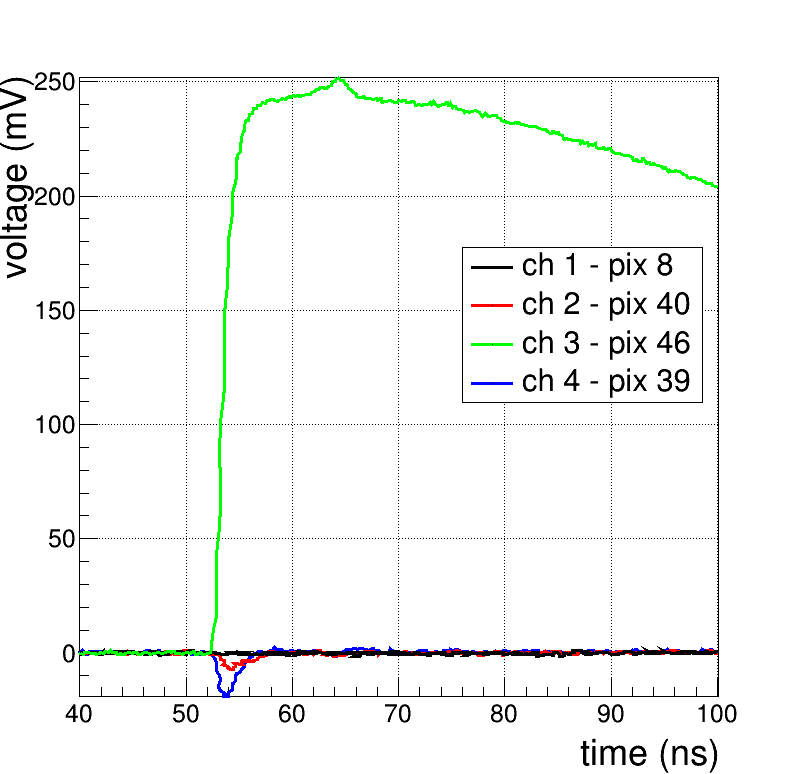}
\caption{}
\label{fig:crosstalk-slow}
\end{subfigure}
\caption{(a) Electrical crosstalk when using the FOUT recorded in pixels  54 and 64, with pixel 61 under illumination, and (b) electrical crosstalk when using the SOUT recorded in pixels 40 and 39, with pixel 46 under illumination.}
\label{fig:crosstalk}
\end{figure}

Before describing the techniques and results of the fully assembled OS-SVSC prototype characterizations, we first note some measurements on crosstalk investigated further based on initial anecdotal observations.  A waveform coming from a crosstalk event shows a negative voltage value when the rise of a positive pulse in a neighbor channel is present for both the FOUT and the standard output (SOUT). Particularly, a damped sine wave as the voltage returns to the baseline is observed for the FOUT. Examples of electrical crosstalk waveforms from the FOUT and SOUT are shown in Fig.~\ref{fig:crosstalk-fast} and Fig.~\ref{fig:crosstalk-slow}, respectively.

To understand these effects, we set up a collimated, pulsed laser (ultrashort pulses down to \SI{50}{ps} at \SI{408}{nm}) attached to a 2D motor stage, incident on the face of one ArrayJ connected to the Ultralytics adapter card. For each acquisition set, a DRS4 evaluation board ~\cite{bib:Bitossi} reads out four pixels at \SI{5.12}{GSa/s}. The laser provides an external trigger and tours each of the 64 pixels. We recorded 1,000 signals per illuminated pixel. We acquired sixteen acquisition sets of four channels to measure the 64 pixels from the ArrayJ. The FOUT and the SOUT were recorded in separate acquisition tests, utilizing the commercial adapter board ArrayX-64 A.3 from Ultralytics for reading the SOUT.
\begin{figure}[!h]
\centering
\begin{subfigure}{.45\textwidth} \centering 
\includegraphics[width=\linewidth]{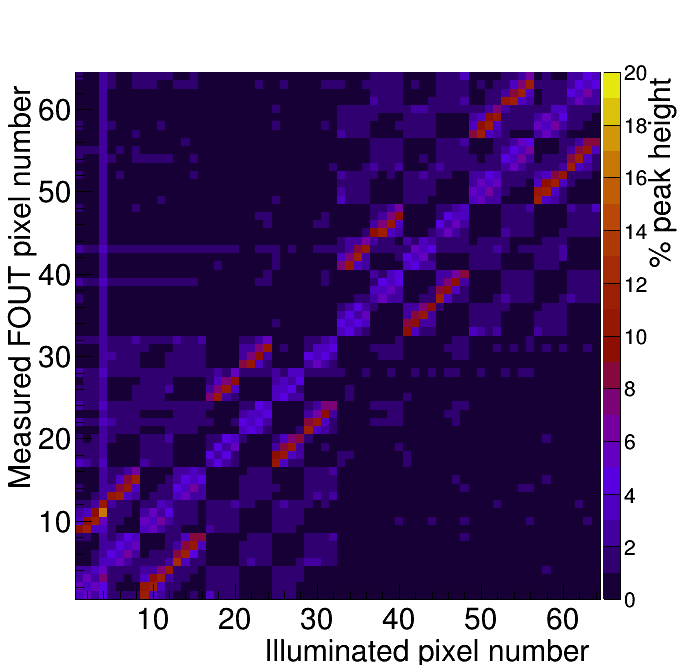}
\caption{}
\label{fig:correlation-fast}
\end{subfigure}
\begin{subfigure}{.45\textwidth} \centering 
\includegraphics[width=\linewidth]{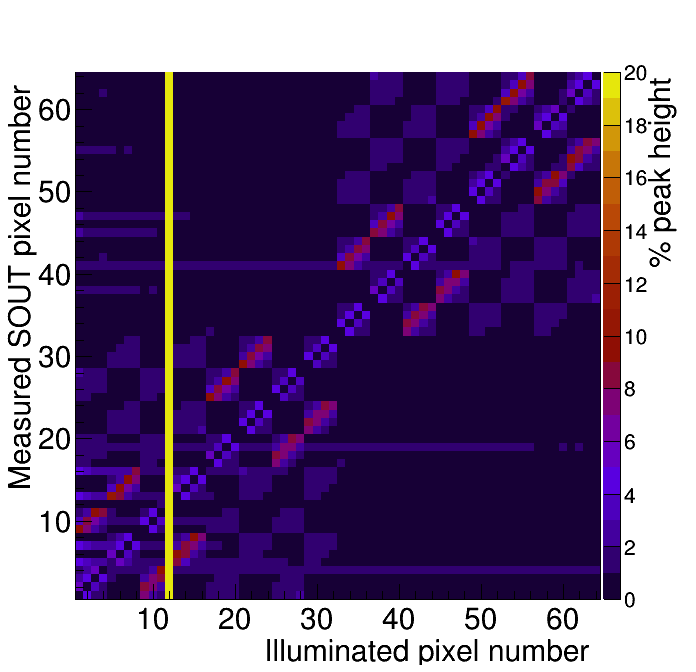}
\caption{}
\label{fig:correlation-slow}
\end{subfigure}
\begin{subfigure}{.7\textwidth} \centering 
\includegraphics[width=\linewidth,angle=90]{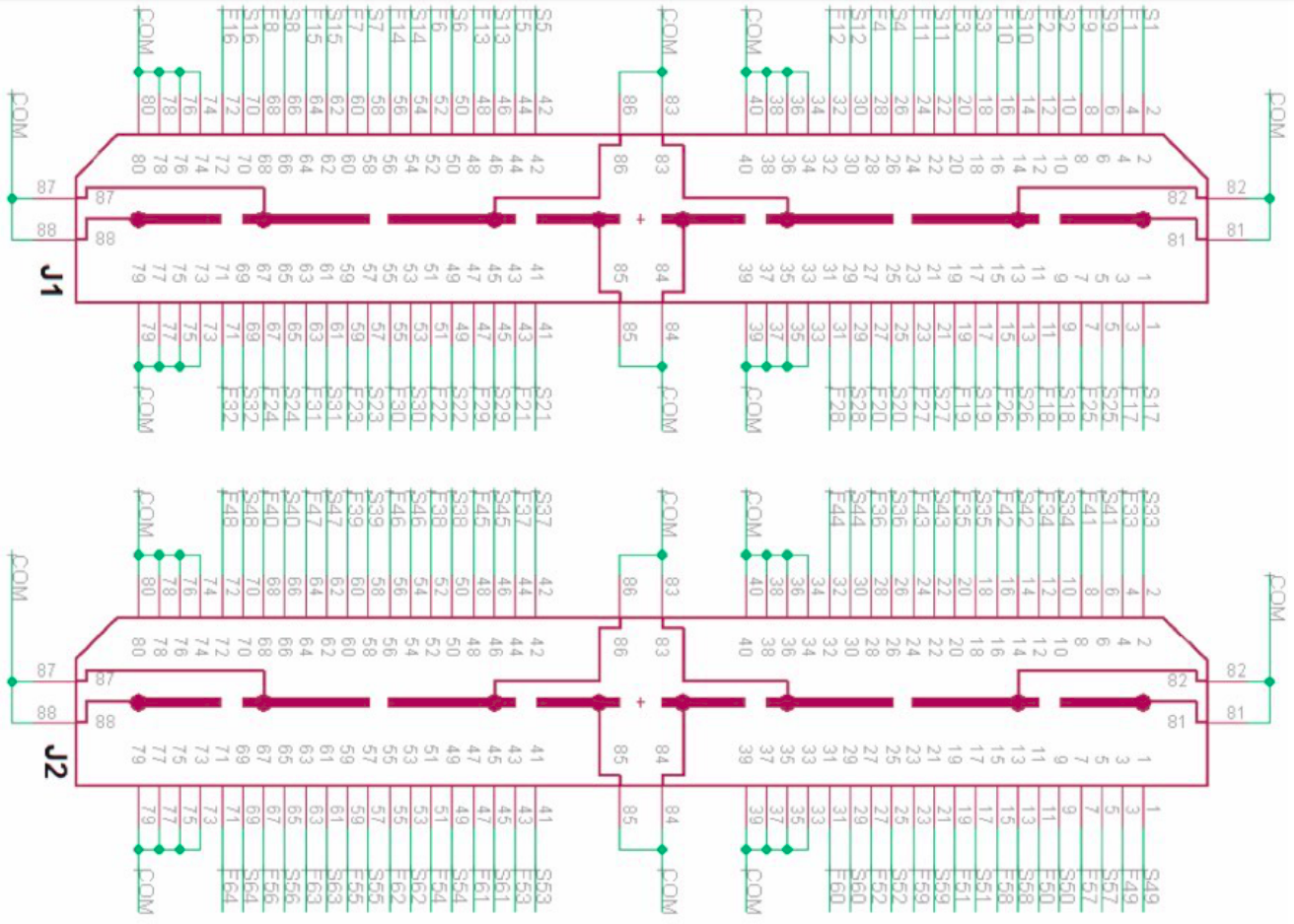}
\caption{}
\label{fig:pinout}
\end{subfigure}
\caption{(a) Correlation matrix from the FOUT and (b) correlation matrix from the SOUT. Both correlation matrices describe the response in all pixels when one of them is illuminated by the laser at a time. The number of each pixel corresponds to the same defined in the ArrayJ datasheet ~\cite{bib:SensL2}. The z-scale corresponds to the percentage of the maximum height of the negative pulse seen in each pixel, with respect to the maximum pulse height of the pixel under illumination. (c) Connector schematic for the ARRAYJ-60035-64P, obtained from~\cite{bib:SensL2}.}
\label{fig:correlation_matrices}
\end{figure}

The maximum pulse height illuminated by the laser and the negative pulse height seen in each pixel are measured to quantify the crosstalk, as characteristic negative pulses are present on crosstalk waveforms for both outputs. The correlation matrices that describe crosstalk in percentage are shown in Figs.~\ref{fig:correlation-fast}, and ~\ref{fig:correlation-slow}, for the FOUT and SOUT, respectively. It should be clarified that the artifacts seen in channels from pixel 4 of the FOUT and pixel 12 of the SOUT are due to damage found in their correspondent Ultralytics cards.

The correlation matrices reveal that the crosstalk is localized in physical groupings of channels in the Samtec QSE connectors used to extract signals from the SensL ArrayJ. Likewise, the matrices show specific groups into quadrants of the ArrayJ. The crosstalk distribution along the pixels for the FOUT and the SOUT are consistent with each other.  When a pixel is illuminated, the impacted channels are the ones with pins located in the same line of the connector, according to the layout schematic of the ArrayJ, shown in Fig.~\ref{fig:pinout}. Furthermore, the highest crosstalk levels measured are located at the two adjacent pins to the illuminated pixel pin. A still detectable crosstalk level is measured on pins that share the same bar ground from the same section in the connector.
Nevertheless, the crosstalk does not extend to independent bar grounds or different connectors. The maximum measured crosstalk reached up to \SI{16.6}{\percent} for the FOUT and up to \SI{10.4}{\percent} for the SOUT. These findings suggest that Samtec connectors in the ArrayJ, and the corresponding mating connectors on the Ultralytics adapter cards could be the dominant contribution to the measured crosstalk.

These observations have multiple implications for the use of the ArrayJ and corresponding ArrayX adapters due to the fact that the FOUTs used in our detector contributes crosstalk signals with positive voltage. The crosstalk is likely to negatively impact our detector's performance, especially in the case where adjacent bars in the same quadrant of the array have an optical signal. Both pulse amplitude and, critically, pulse timing estimates will be degraded for both signals in this case. More broadly, we note that these observations have motivated our collaboration to explore making custom arrays for future work, rather than using the commercially available arrays.

\section{Electronics Calibration and Waveform Processing}
In this section, we describe the offset and timing calibrations of the 
electronics, and the waveform processing techniques applied to the 
calibrated waveforms.
\subsection{Electronics Calibrations}
\label{sec:elec_calib}
\begin{figure}[!h]
\begin{subfigure}{.50\linewidth}
\includegraphics[width=\linewidth]{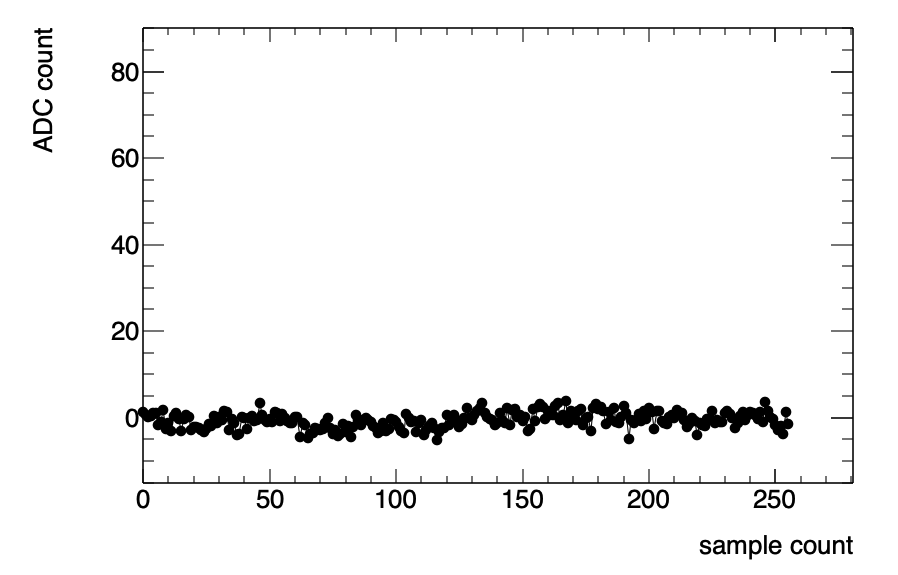}
\caption{}
\label{fig:ex_ped_sub1}
\end{subfigure}%
\begin{subfigure}{.50\linewidth}
\includegraphics[width=\linewidth]{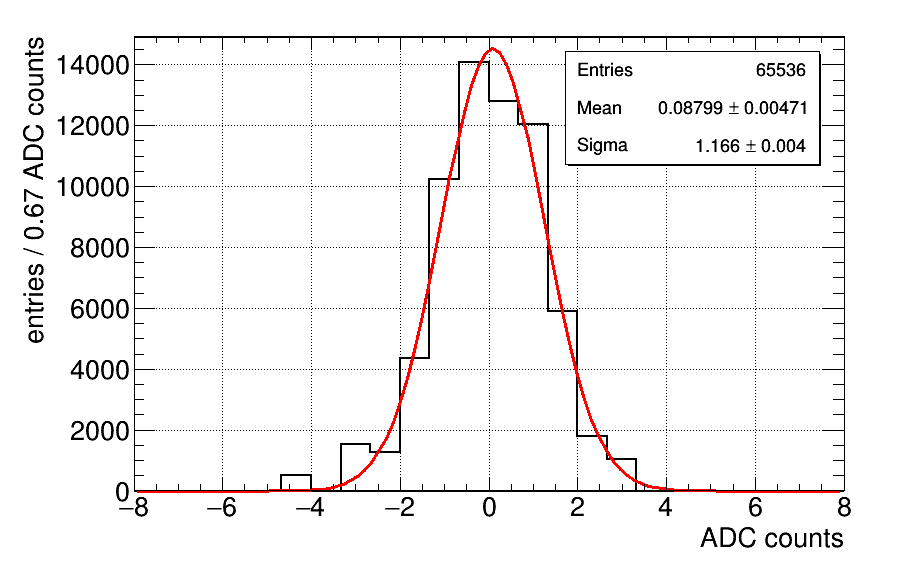}
\caption{}
\label{fig:ex_ped_sub2}
\end{subfigure}
\caption{(a) Example pedestal-subtracted waveform with no input signal. 
(b) ADC count distribution for the noise waveform.} 
\label{fig:pedestal_examples}
\end{figure}

Fabrication variations from the IRS ASIC individual capacitors and comparators generate sample-to-sample differences creating a unique offset voltage value on each sample. The subtraction of these values allows for acquiring proper waveforms. The pedestal value is determined as the average value of the measurement when random sampling is performed with no external input. An example pedestal subtracted noise trace and corresponding noise histogram is shown in Fig.~\ref{fig:pedestal_examples}. Typical noise values for pedestal-subtracted traces range from \SIrange{1.5}{3}{ADC}, corresponding to $\sim$\SIrange{1}{2}{mV}.

\begin{figure}[!h]
\centering
\includegraphics[width=0.60\textwidth]{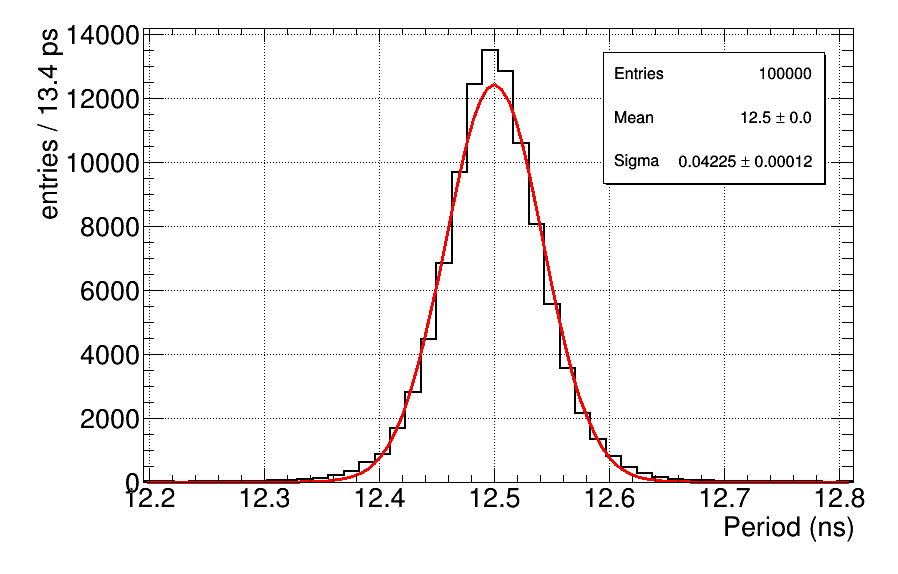}
\caption{Example distribution of measurements of the pulse period of calibration signals on a channel after timing calibration is performed. The injected pulse separation is \SI{12.5}{ns}, matching the post-calibration value. The channel shows a timing resolution of $\sigma = \SI{42}{ps}$.}
\label{timing_calib}
\end{figure}

The ASICs at the core of the readout electronics utilize delay lines to toggle an array of switched capacitors. 
Process variations in the fabrication of individual transistors along the delay line result in varying timing separations from sample to sample, which must be adequately calibrated to achieve timing resolutions on the order of \SI{100}{ps} or better~\cite{bib:Varner}. A matrix-based time-period calibration method was utilized to determine the sampling delays of the 32,768 storage cells (2 ASICs $\times$ 128 sampling delays $\times$ 128 channels)~\cite{bib:Cheng}. Two input pulses were sent through an SMA calibration input on each carrier with a known delay time. The pulse rise times were chosen to roughly match those seen in a typical SiPM pulse.  The precision of the solution was verified by calculating the period of the input pulse with the calibrated timing values for each sampling cell.  An example of the distribution of the pulse period measurement is shown in Fig.~\ref{timing_calib}.

\begin{figure}[!h]
\begin{subfigure}{.50\linewidth}
\includegraphics[width=\linewidth]{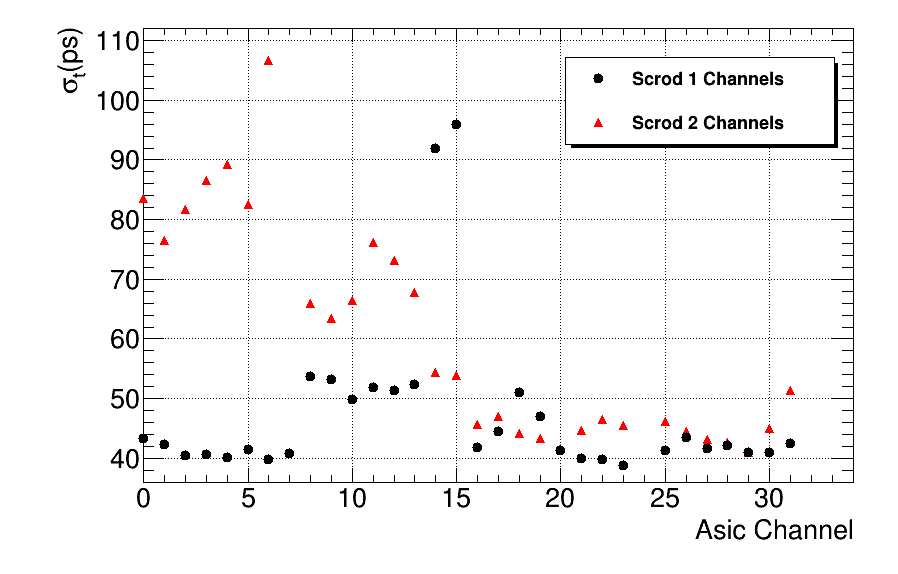}
\caption{}
\label{fig:time_calibration1}
\end{subfigure}%
\begin{subfigure}{.50\linewidth}
\centering
\includegraphics[width=\linewidth]{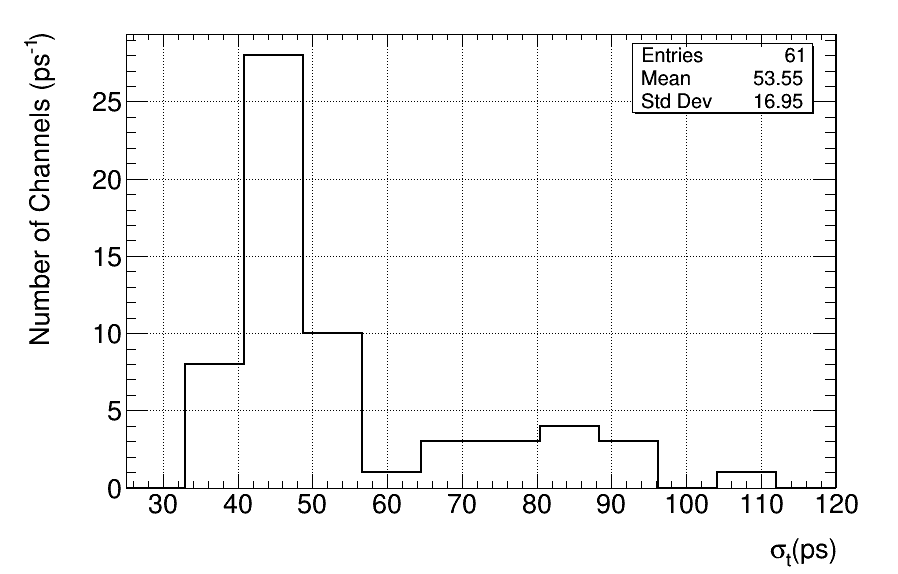}
\caption{}
\label{fig:time_calibration2}
\end{subfigure}
\caption{(a) Timing Calibration values vs. channel number, (b) Timing calibration value distribution.}
\label{fig:time_calibration}
\end{figure}

An external ROOT file stores the calibration values along with the output data during the acquisition process. Fig.~\ref{fig:time_calibration} summarizes the resulting timing resolutions from the characterized channels. The average timing resolution for each channel is $\sigma \approx$\SIrange{30}{90}{ps}.  Variations are attributed in part to jitter introduced from the FPGA and PCB in the distribution of timing strobes. In addition, a number of internal timing parameters must be tuned for each ASIC in order to avoid timing overlaps or gaps at the wraparound locations of the sampling array. Note that the timing resolutions here are all based on pairs of identical pulses measured within a single channel, they do not represent expected coincidence time resolutions between channels measuring a wider variety of pulse shapes.

\subsection{Waveform Processing}
\begin{figure}[!h]
\begin{subfigure}{.50\linewidth}
\includegraphics[width=\linewidth]{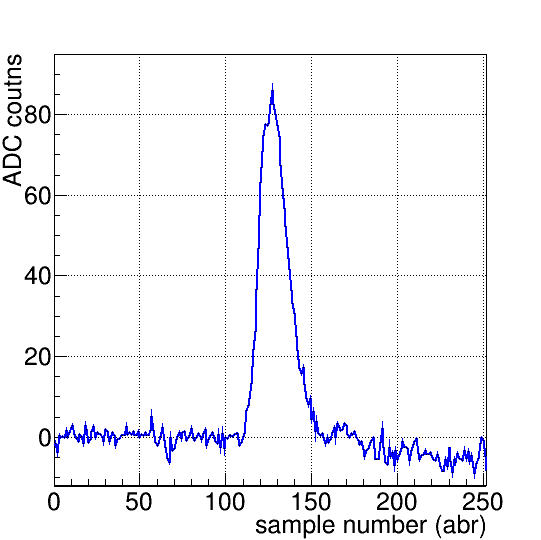}
\caption{}
\label{fig:test1_sub1}
\end{subfigure}%
\begin{subfigure}{.50\linewidth}
\centering
\includegraphics[width=\linewidth]{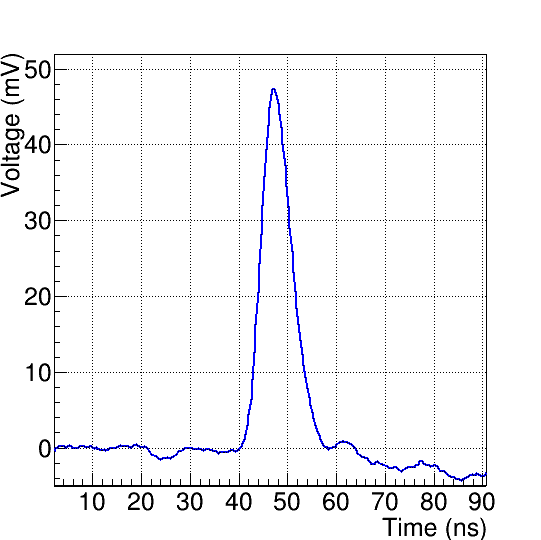}
\caption{}
\label{fig:test1_sub2}
\end{subfigure}
\caption{Example of the waveform processing. (a) is a raw waveform coming from the digitizer, and (b) corresponds to the same waveform after the signal processing.}
\label{fig:signals}
\end{figure}

The waveforms recorded from the IRS3D have 256 samples for the particle emission sources tests and 384 samples for muon acquisitions. In both cases, the acquisition was performed at \SI{2.73}{GSa/s} and the FOUTs of the ArrayJs were stored. Any residual baseline offset is eliminated by calculating the mean of the signal base using the first 64 samples and subtracting that value from all the waveform samples. A 5-sample moving average filter was applied to smooth the signal. In some cases, ASIC waveforms exhibited artifacts with two or more spikes, while other waveforms showed significant electrical crosstalk, as discussed in Section~\ref{sec:crosstalk}. About \SI{12.7}{\percent} of events were discarded due to these defects. ROOT-based analysis software was used to process the pulses. The time scale is obtained from the sample sequence on each waveform using the time calibration values described in Section~\ref{sec:elec_calib}. The amplitude and the leading edge time at 50\% of the maximum were measured for each waveform. An example of a waveform before and after processing is shown in Fig.~\ref{fig:signals}. 

\section{Detector Calibration}
\label{sec:calibration_section}
Previous work~\cite{bib:Sweany} employed $^{137}$Cs and $^{22}$Na sources for energy calibrations, and either collimated $^{90}$Sr source or the back-to-back gammas from a $^{22}$Na source to scan each bar along its length and generate a calibrated response of amplitude- and time-based functions from SiPM waveforms taken at each end of a bar. However, in the OS-SVSC design it is not possible to apply this technique for each bar, because the 28 outer bars ``shield'' the remaining 36 bars, preventing particles from the calibration source to reach the inner bars without scattering in the outer bars.

In order to calibrate the position in the inner bars, we have developed a calibration technique for the OS-SVSC taking advantage of cosmic ray muons, which are the most abundant naturally available charged particles at sea level~\cite{bib:Cecchini}. This method utilizes the outer bar position calibrations to construct a ``muon telescope''. We use the calibration from $^{90}$Sr scans of the outer bars to reconstruct the muon trajectory, then project it through the array to calibrate the inner bars.

With the condition that all channels reading the bars from the same column must pass a threshold of \SI{30}{ADC} (\SI{\sim 18.31}{mV}) to acquire a muon event, we expect to observe only paths of muons running through a full column of bars. The geometry of the OS-SVSC, combined with the trigger condition, allows us to record muons with angles of up to \ang{76.15} with respect to the zenith. However, the cosmic muon intensity follows a $\cos^{2}\theta$ distribution, where $\theta$ is the zenith angle: therefore, a small number of muon events will have an angle higher than \ang{60} with respect to the zenith. 

\begin{figure}[!h]
\centering
\includegraphics[width=0.52\textwidth]{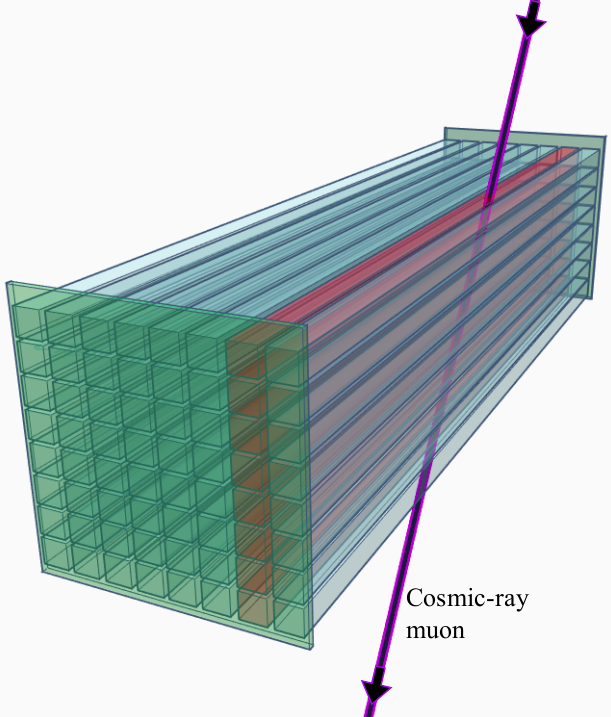}
\caption{Illustration of a muon passing through a column of bars. The trigger is
activated when a pulse on each channel from a vertical column of bars passes a
threshold. The triggered bars are marked in red color.}
\label{fig:muon}
\end{figure}

Muon data can also be used to calibrate energy. Muons have an average energy of approximately 4 GeV at level sea and deposit \SI{\sim 2}{MeV/cm} in the scintillator~\cite{bib:Cecchini}. The advantage of the trigger condition allows us to assure that the recorded energy comes from the muon passing through the scintillator and not from muon decays in the inner bars, so as to not add spurious energy to the measurements. In addition, as these muons travel at almost the speed of light, we can utilize time differences among pairs of bars to obtain time resolutions.

This section describes the techniques for calibrating bars for energy, position, and time using muons, along with the traditional energy and position calibrations using particle emission sources. The tested column of bars is located at the edge of the detector in order to compare muon calibrations with particle source calibrations, and includes bars numbered 8 (top), 16, 24, 32, 40, 48, 56, and 64 (bottom). Unfortunately, the electronics channel that reads bar 8 was not functional. Hence, the calibrations are reported with a primary focus on using seven of the eight available bars.   

\subsection{Energy Calibrations}
\begin{figure}[!h]
\centering
\includegraphics[width=0.6\textwidth]{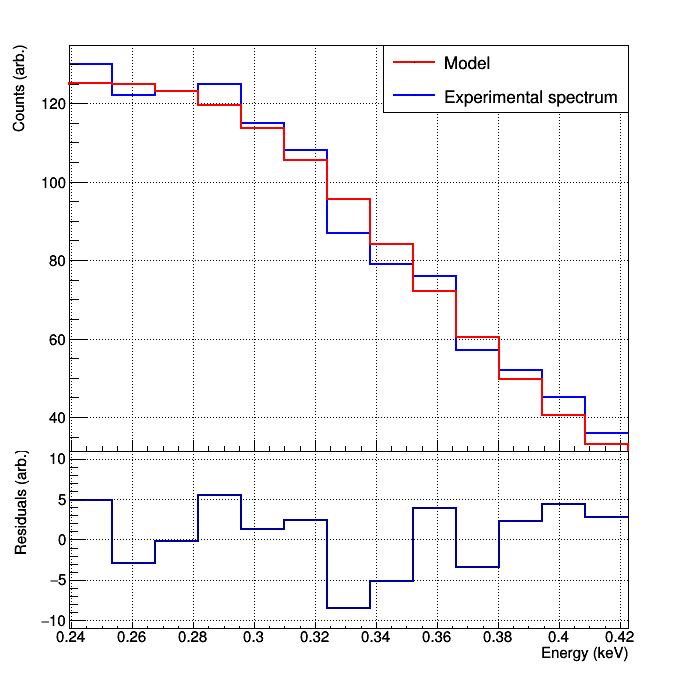}
\caption{Example of an energy calibration. The expected model of a $^{22}$Na gamma-ray source in the Compton edge region \SI{340.67}{keV} fits the amplitude spectrum from bar 56. Residuals are included.}
\label{fig:ecal_example}
\end{figure}

We evaluate and compare two techniques for energy calibration. The first is a technique where the geometric mean of the pulse heights on each end of the given bar is measured and fit to a predicted spectrum model of a $^{22}$Na gamma-ray source around a Compton edge region. The technique is described in more detail in~\cite{bib:Sweany} and is readily available for bars on the edge of the array. The threshold for these acquisitions is \SI{20}{ADC} (\SI{\sim 12.2}{mV}). An example fit to such an amplitude spectrum is shown in Fig.~\ref{fig:ecal_example}. 

As an alternative approach that does not require a source and can easily be 
expanded to any bar of the array, we simulated spectra obtained from cosmic ray 
muons and determine the most probable value (MPV) of their peaks.
Simulations of the OS-SVSC were performed in Geant4~\cite{bib:Geant4} 
version 10.04. 
The input of $\sim$6,000 secondary muons randomly distributed on an area of \SI{5}{mm} $\times$ \SI{350}{mm} above the top bar, allows muons with large zenith angles to cross the bar set. The input was generated in CORSIKA~\cite{bib:CORSIKA} version 73900 from primary protons with zenith angles from 0$^{\circ}$ to 70$^{\circ}$, azimuth angle set at 0$^{\circ}$, a slope of the energy spectrum of $-2$, and an observation altitude at sea level.

\begin{table}[!h]
\caption{Comparison of MPVs for real and simulated muon data, using the energy calibrations derived from the $^{22}$Na Compton edge for a column of bars on the edge of the detector.}
\begin{center}
\begin{tabular}{C{1.5cm}C{4.0cm}C{3.0cm}C{2.5cm}}
\hline
\textbf{Bar number} & \textbf{Energy conversion using Compton edge (keVee/mV)} &  \multicolumn{2}{l}{\textbf{Muon energy deposition MPV (keVee)}} \\
\hline 
 &  & Data & Simulations\\
\hline
16  &  20.50   &   \num{968.6 \pm 2.8}   &   \num{1015 \pm 6.8}   \\
24  &  20.51   &   \num{984.4 \pm 2.9}   &   \num{1026 \pm 6.2}   \\
32  &  24.0    &   \num{985.8 \pm 2.9}   &   \num{1020 \pm 6.6}   \\
40  &  19.57   &   \num{986.7 \pm 3.1}   &   \num{1026 \pm 6.5}   \\
48  &  18.04   &   \num{987.1 \pm 2.8}   &   \num{1018 \pm 6.2}   \\
56  &  18.43   &   \num{985.6 \pm 2.7}   &   \num{1023 \pm 5.7}   \\
64  &  19.35   &   \num{929.5 \pm 2.8}   &   \num{1010 \pm 5.1}   \\
\hline
\end{tabular}
\label{tab:energy_cal}
\end{center}
\end{table}

In Table~\ref{tab:energy_cal}, we use the energy conversion obtained from the $^{22}$Na source measurements to compare MPVs for measured and simulated muon data in the set of edge bars along the array. The energy conversion is used to translate the MPVs of observed cosmic ray distributions from ADC counts into energy units. These measurements present lower values than those from the muon simulation. The measured MPVs are different from each other by up to \SI{5.83}{\percent} and the differences with the corresponding simulated MPV ranged from \SI{3}{\percent} up to \SI{7.97}{\percent}. By knowing these differences, the MPV from the real muon data can be used to determine energy conversion values for the rest of the bars.

\subsection{Position Response}
The interaction position along a bar can be determined by using either the time or amplitude measurements from the bar's ends, i.e., using the functions $t_{1}-t_{2}$ or $\ln \frac{A_{1}}{A_{2}}$, where $t_{1}$ and $t_{2}$ are the arrival times of the pulses and $A_{1}$ and $A_{2}$ are the pulse heights measured on the two ends of the bar. A detailed description of both approaches can be found in~\cite{bib:Sweany}.

\subsubsection{Position calibration using a particle source}
\label{sec:pos_part_source}
The experimental setup consists of a lead collimated $\beta$-emitting $^{90}$Sr source, supported by a bracket mounted to a 2D Velmex XSlide (XN10-0120-M01-71) and driven by two stepper motors (PK245-01AA). The lead collimator limits the range of incident positions of betas within 2 mm on the bars. The source position, controlled via USB, was moved along each target bar length in steps of \SI{5}{mm}; 5,000 events were acquired per source position.

\begin{figure}[!h]
\begin{subfigure}{.45\linewidth}
\includegraphics[width=\linewidth]{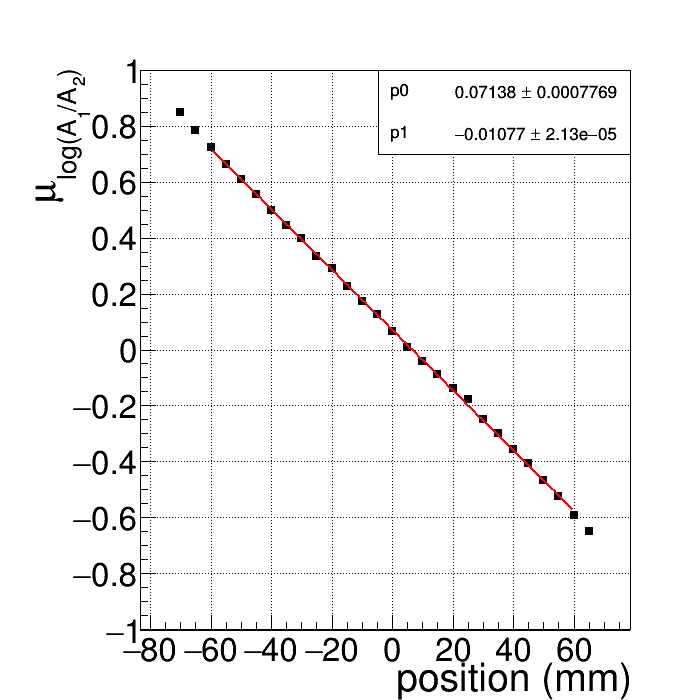}
\caption{}
\label{fig:test3_sub1}
\end{subfigure}%
\begin{subfigure}{.45\linewidth}
\centering
\includegraphics[width=\linewidth]{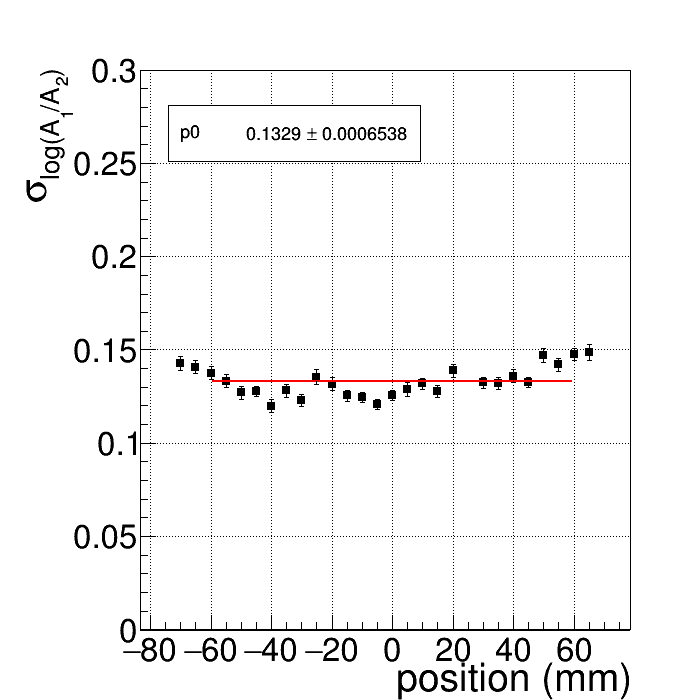}
\caption{}
\label{fig:test3_sub2}
\end{subfigure}
\caption{Calibration example for the amplitude-based position resolution 
measurement, bar 40. (a) The logarithm of the ratio of the amplitudes as a function 
of the source position and (b) the standard deviation of the logarithm 
of the ratio of the amplitudes as a function of the source position. Only depositions from \SIrange{900}{1000}{keVee} are included.}
\label{fig:amplitude-slopes}
\end{figure}

\begin{table}[!h]
\caption{Summary of the position resolution results when using a $^{90}$Sr source. Only depositions from \SIrange{900}{1000}{keVee} are considered. The errors represent the statistical uncertainty from the fit of the particular distribution.}
\begin{center}
\begin{tabular}{c c c c}
\hline
Bar ID	&$\sigma_z^{A}$ (mm)		&$\sigma_z^{t}$ (mm) 			& $\sigma_z$ (mm)		 \\	
\hline
16	& \num{15.67 \pm 0.08}	& \num{37.49 \pm 0.34} & \num{14.46 \pm 0.10}\\
24	& \num{14.32 \pm 0.07}	& \num{43.58 \pm 0.38} & \num{13.60 \pm 0.11}\\
32	& \num{16.25 \pm 0.13}	& \num{21.99 \pm 0.25} & \num{13.07 \pm 0.05}\\ 
40	& \num{12.35 \pm 0.06}	& \num{41.41 \pm 0.45} & \num{11.83 \pm 0.12}\\ 
48	& \num{8.34 \pm 0.06}	& \num{31.56 \pm 0.59} & \num{8.06 \pm 0.14}\\ 
56	& \num{8.91 \pm 0.07}	& \num{33.28 \pm 0.56} & \num{8.61 \pm 0.14}\\
64	& \num{8.63 \pm 0.07}  	& \num{27.09 \pm 0.41} & \num{8.23 \pm 0.11}\\
\hline  
\end{tabular}
\label{tab:posRes_sources}
\end{center}
\end{table}
The distributions of $t_{1}-t_{2}$ and $\ln \frac{A_{1}}{A_{2}}$ were fit with Gaussian functions for every source position. Plots of the mean and sigma values as a function of the position of the source are used to obtain the position resolution $\sigma_{z}$, calculated as $\frac{\sigma_{\text{avg}}}{p_{1}}$, where $\sigma_{\text{avg}}$ is the average of the standard deviation as a function of $z$ and $p_{1}$ is the slope of the first-order polynomial fit to the mean as a function of $z$. An example of amplitude-based position calibrations plots for one of the tested bars is shown in Fig.~\ref{fig:amplitude-slopes}. An error of 1.44 mm contributes to the position resolution measurements due to the collimator width.  Table~\ref{tab:posRes_sources} summarizes position resolutions when using amplitude-based measurements, time-based measurements and their combination using the best linear unbiased estimator (BLUE)~\cite{bib:Lista}. 
\subsubsection{Position Calibration using muons}
\label{sec:pos_part_muon}

\begin{figure}[!h]
\begin{subfigure}{\linewidth}
\centering
\includegraphics[width=\linewidth]{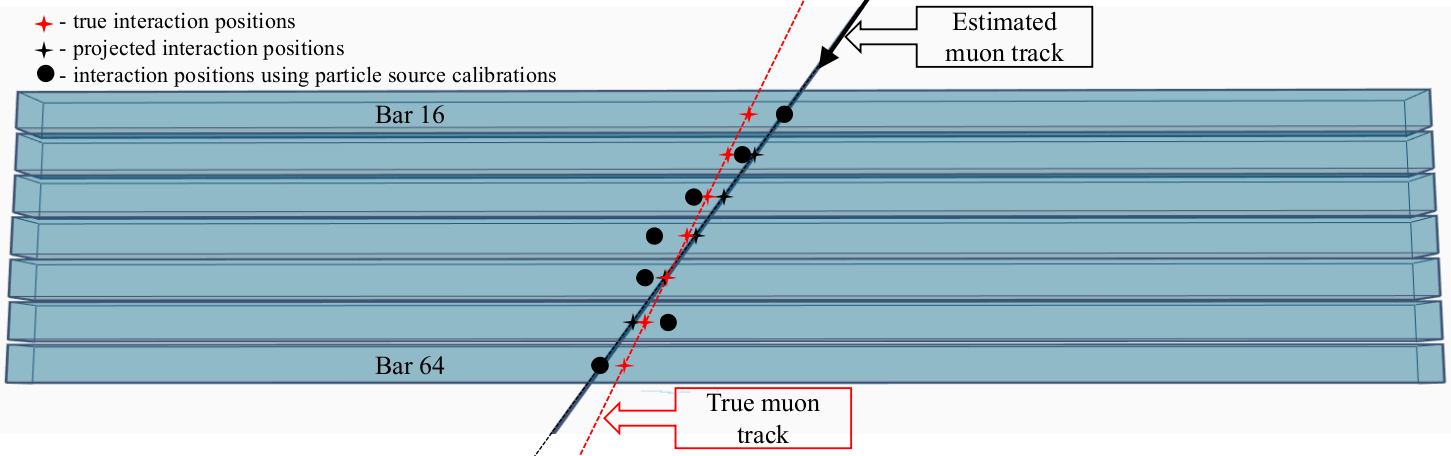}
\caption{}
\label{fig:calibrationGeometry}
\end{subfigure}
\begin{subfigure}{.5\linewidth}
\centering
\includegraphics[width=0.9\linewidth]{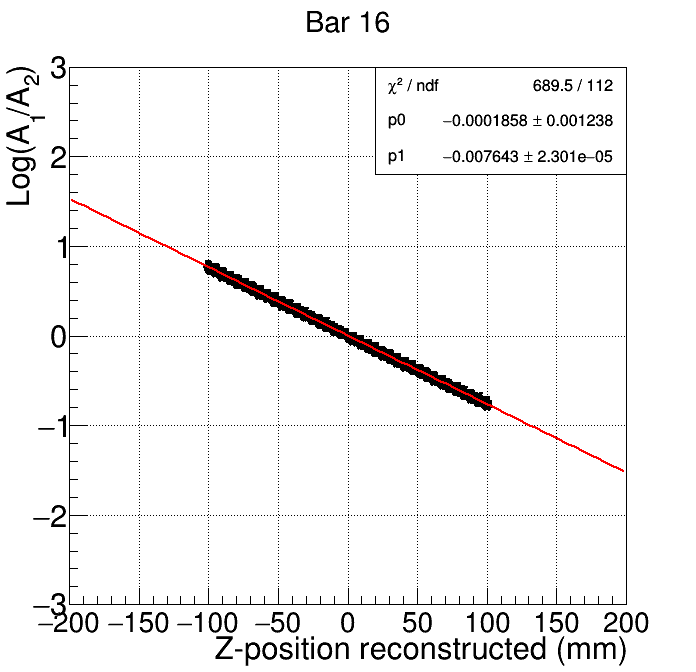}
\caption{}
\label{fig:bar16}
\end{subfigure}%
\begin{subfigure}{.5\linewidth}
\centering
\includegraphics[width=0.9\linewidth]{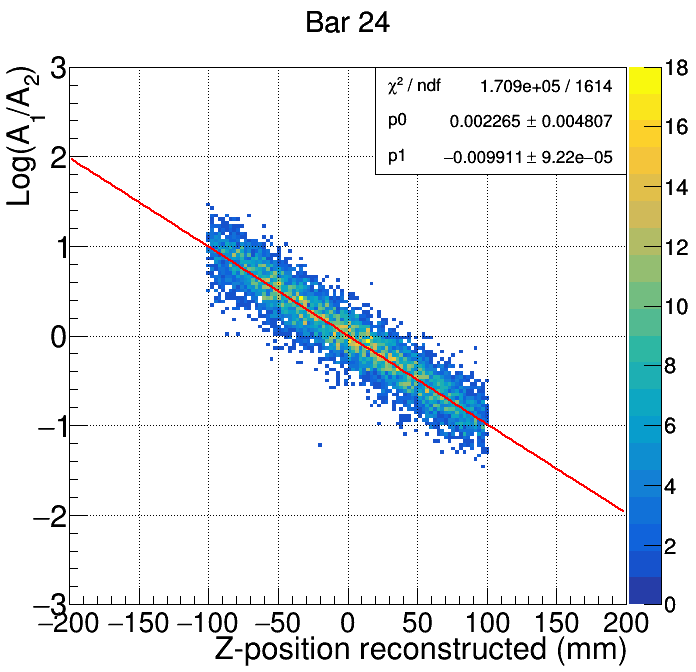}
\caption{}
\label{fig:bar24}
\end{subfigure}
\caption{(a) Illustration of a muon passing through the studied bar set, (b) relation between the logarithm of the ratio of the amplitudes vs. the reconstructed interaction position for the top bar (16) using the $p_{1}$ slope value, (c) relation between the logarithm of the ratio of the amplitudes vs. the estimated interaction position for an inner bar (24).}
\label{fig:muonPositionCalibration}
\end{figure}

\begin{figure}[!h]
\begin{subfigure}{.45\linewidth}
\includegraphics[width=\linewidth]{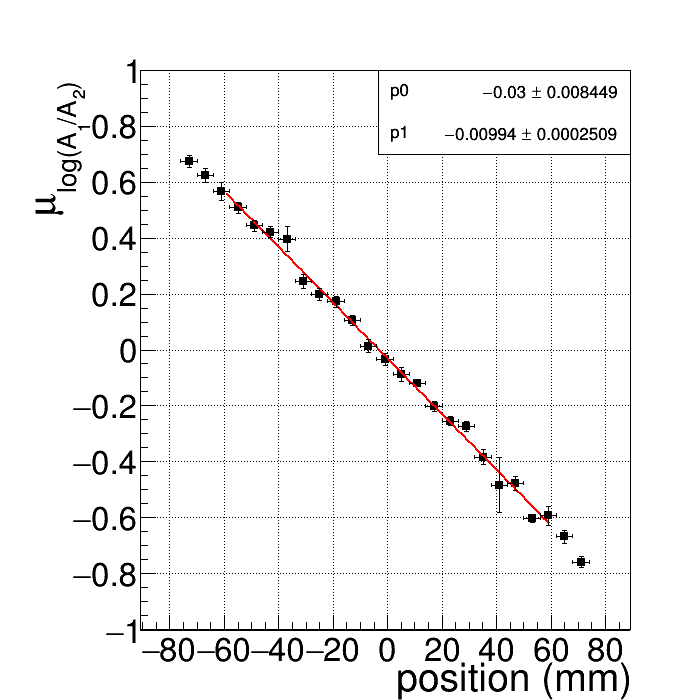}
\caption{}
\label{fig:posResolutionAmplitudes_mean}
\end{subfigure}%
\begin{subfigure}{.45\linewidth}
\centering
\includegraphics[width=\linewidth]{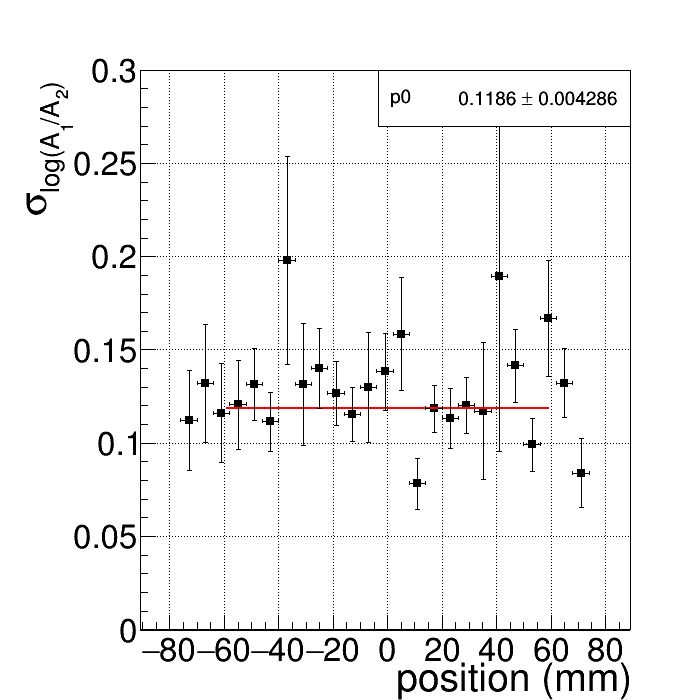}
\caption{}
\label{fig:posResolutionAmplitudes_std_dev}
\end{subfigure}
\caption{(a) The logarithm of the ratio of the amplitudes as a function of the probable muon $z$-position and (b) the standard deviation of the logarithm of the ratio of the amplitudes as a function of the probable muon $z$-position for bar 40. Only depositions from \SIrange{900}{1000}{keVee} are included.}
\label{fig:posResolutionAmplitudes}
\end{figure}

We tested the muon calibration technique with the same column of edge bars calibrated with particle emission sources. Only amplitude-based measurements for the $\sim$8,000 events were recorded. The mean values of the logarithm of the ratio of the amplitudes distributions were moved to zero for alignment among all the bars. For each event, the interaction positions from the top and bottom bars along the $z$ axis create the muon trajectory, as illustrated in Fig.~\ref{fig:calibrationGeometry}. These two interaction positions are obtained using the logarithm of the ratio of the amplitudes measured from muon events and the $p_{1}$ slope value of the calibrations using particle sources, as shown in Fig.~\ref{fig:bar16}. 
The muon trajectory is then projected to the rest of the bars. We directly associated the interaction positions of the inner bars with the corresponding amplitude measurements. An example of a 2D distribution representing such an association for bar 24 is shown in Fig.~\ref{fig:bar24}. Each 2D distribution is then binned along the interaction position axis (with a bin width of \SI{6}{mm}). Every bin is projected along the log amplitude ratio axis to generate sets of 1D distributions. The mean and standard deviation values from Gaussian fits of the distributions allowed to determine the position resolution via the same approach described in Section~\ref{sec:pos_part_source}. An example for depositions from \SIrange{900}{1000}{keVee} is shown in Fig.~\ref{fig:posResolutionAmplitudes}.

The reconstructed interaction positions from the top and bottom bars have associated errors from the previous calibrations, propagated to the muon trajectory and, consequently, to the reconstructed positions of the inner bars.
We implemented an algorithm based on the Monte Carlo (MC) method to approximate the uncertainty due to the error in the estimated muon trajectory. Simulated true muon trajectories from the top and bottom bars' positions were randomly generated following the expected angular cosmic-ray muon distribution. A second track, which represents the measured muon path, is generated by smearing the interaction positions from the top and bottom bars, using the actual amplitude-based position resolution values reported in Table \ref{tab:posRes_sources}. For each trial, the true and the estimated muon trajectories were projected to the rest of the bars. 
\begin{figure}[!ht]
\centering
\includegraphics[width=0.6\linewidth]{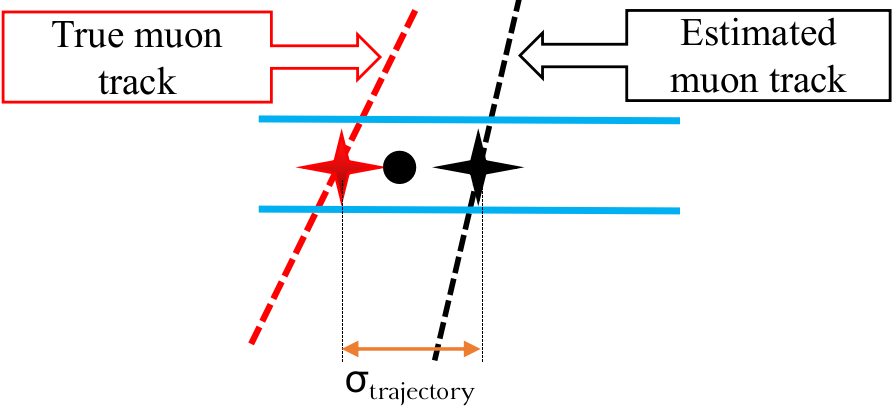}
\caption{Illustration of $\sigma_{\text{trajectory}}$. The red star represents the true interaction position. The black star represents the estimated interaction position from the muon trajectory generated with smeared top and bottom bars interaction positions. The black dot represents a measurement obtained during this event.}
\label{fig:MC_diagram}
\end{figure}
Fig.~\ref{fig:MC_diagram} shows an example of an MC trial for an individual bar. $\sigma_{\text{trajectory}}$ represents the uncertainty of the estimated muon track due to the errors present in the top and bottom bars interaction positions. Values of $\sigma_{\text{trajectory}}$ are obtained for each bar using 100,000 MC trials fitted with Gaussian functions.  
The width of the Gaussian fit values, along with the amplitude-based position resolutions using muons, are shown for each bar in Table~\ref{tab:posRes_muons}.

\begin{table}[!ht]
\caption{Summary of the position resolutions when calibrating the inner bars with muon data $\sigma_z^{A_{\mu}}$ and the MC errors for inner bars with ``true'' position resolutions of \SI{15}{mm}. Only depositions from \SIrange{900}{1000}{keVee} are considered in the real data. The errors on the resolutions are statistical.}
\begin{center}
\begin{tabular}{c c c }
\hline
Bar ID	&$\sigma_z^{A_{\mu}}$ (mm)	&	MC$\sigma_{trajectory}$	 \\	
\hline
24	& \num{15.63 \pm 0.79}	 &	\num{13.16 \pm 0.03}			\\
32	& \num{15.98 \pm 0.70}	 &	\num{10.87 \pm 0.02}		 	\\ 
40	& \num{11.93 \pm 0.52} 	 &	\num{8.97 \pm 0.02}			 	\\ 
48	& \num{13.02 \pm 0.56} 	 &	\num{7.81 \pm 0.02}	    	\\ 
56	& \num{17.74 \pm 0.96}	 &	\num{7.65  \pm 0.02}	        \\
\hline  
\end{tabular}
\label{tab:posRes_muons}
\end{center}
\end{table}

\subsection{Time Resolution}
\label{sec:time_resolution}
The same muon data used for position calibrations were utilized to obtain the time resolution among the bars and to determine time deviations. The measured time difference between the interactions occurring in two bars, $\Delta T_{i,\text{ref}}$, can be defined as:
\begin{equation}
\Delta T_{i,\text{ref}} = \frac{(t_{1}+t_{2})_{\text{ref}}}{2} - \frac{(t_{1}+t_{2})_{i}}{2} + \delta_{i,\text{ref}}
\end{equation}
where $\frac{(t_{1}+t_{2})_{\text{ref}}}{2}$ is the average time of the reference bar, $\frac{(t_{1}+t_{2})_{i}}{2}$ is the average time of the $i$th bar, and $\delta_{i,\text{ref}}$ is the time difference between the $i$th bar and the reference bar due to electronics and physical factors, like cable lengths, pixel time responses, and ASIC fabrication effects, among others. The expression is similar to the one reported in \cite{bib:Zhu}. Nevertheless, our measurements only report on time, no geometric information related to the interaction position along the bars or the distance between the bars was considered.

\begin{figure}[!ht]
\centering
\begin{subfigure}{0.49\linewidth}
\centering
\includegraphics[width=\linewidth]{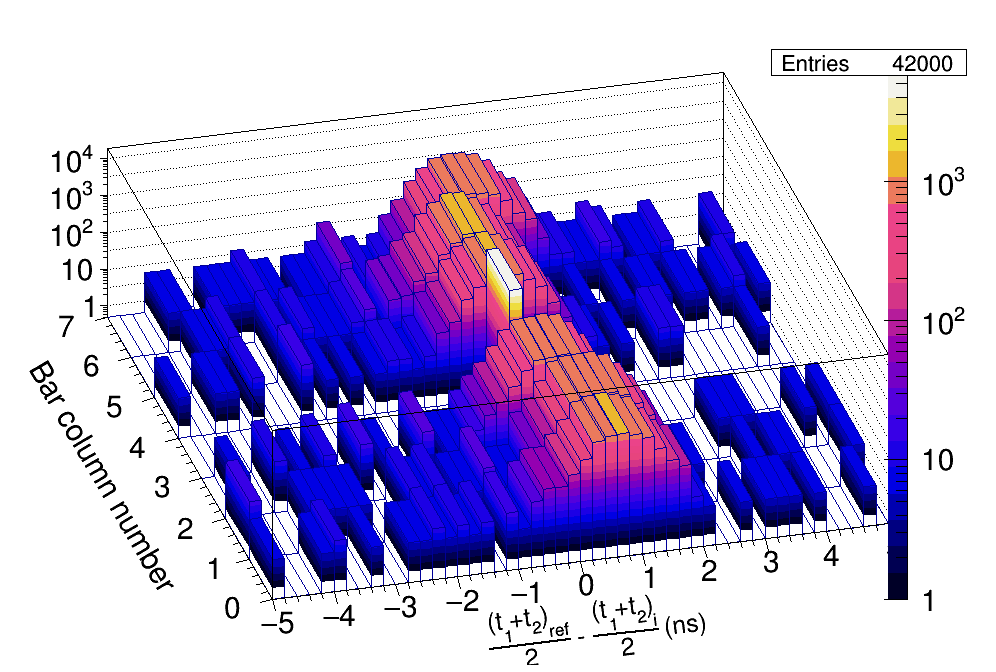}
\caption{}
\label{fig:timeCalibrationBefore}
\end{subfigure}
\begin{subfigure}{.49\linewidth}
\includegraphics[width=\linewidth]{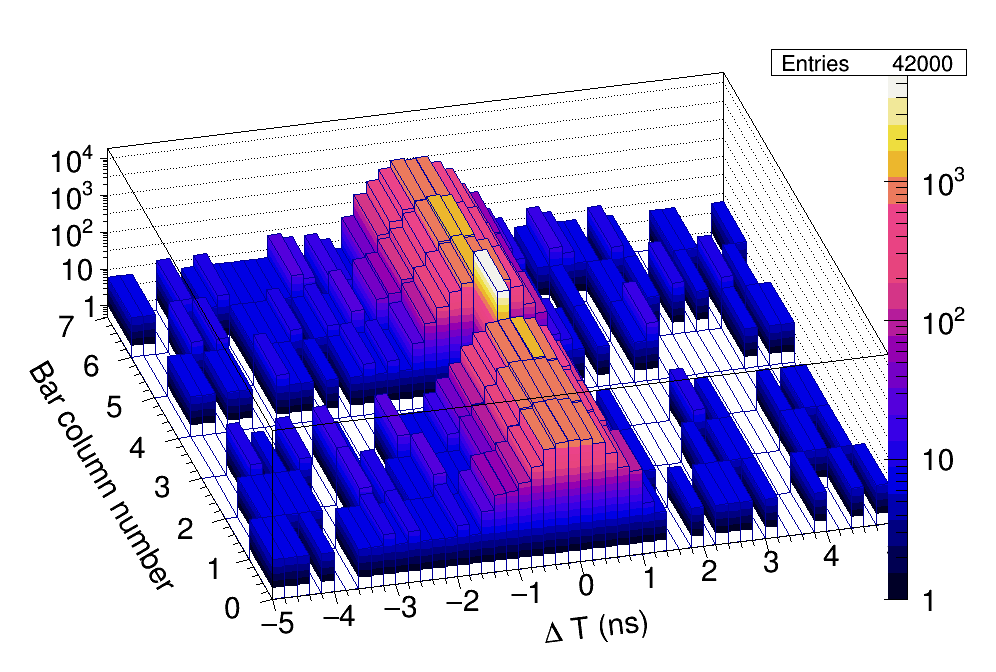}
\caption{}
\label{fig:timeCalibrationAfter}
\end{subfigure}
\begin{subfigure}{.50\linewidth}
\centering
\includegraphics[width=\linewidth]{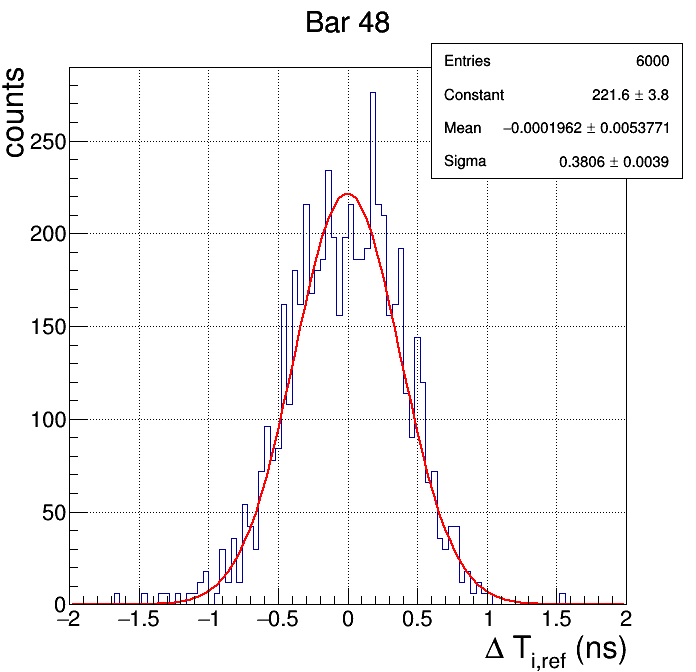}
\caption{}
\label{fig:time_res_bar_48}
\end{subfigure}
\caption{(a) Distributions of $\frac{(t_{1}+t_{2})_{\text{ref}}}{2} - \frac{(t_{1}+t_{2})_{i}}{2}$ for the studied bar set. (b)  Distributions of $\Delta T_{i,\text{ref}}$ for the studied bar set. (c) Distribution of $\Delta T_{i,\text{ref}}$ for bar 48.}
\end{figure}

Bar 40 was defined as the reference bar, hence $\Delta T_{40,40} = 0$. The assignment for a reference bar was made with no specific preference on the placement or the performance of such a bar. The non-centered positions of the peaks for the $\frac{(t_{1}+t_{2})_{\text{ref}}}{2} - \frac{(t_{1}+t_{2})_{i}}{2}$ distributions, shown in Fig.~\ref{fig:timeCalibrationBefore}, suggest evidence of the delay time $\delta_{i,\text{ref}}$ due to physical factors among the pairs of bars. 
It is possible to inter-calibrate bars for the same column when aligning those peaks to zero, thus obtaining $\delta_{i,\text{ref}}$  as a relative time offset. The alignment is shown in Figs.~\ref{fig:timeCalibrationBefore} and~\ref{fig:timeCalibrationAfter}. 
The standard deviation of a Gaussian function fit from $\Delta T_{i,\text{ref}}$ represents the time resolution, as is shown in Fig~\ref{fig:time_res_bar_48} for bar 48. The time resolution results, summarized in Table~\ref{tab:time_resolution}, show values of about \SI{400}{ps}, with variations among the channels of \SI{10}{\percent} or less.

\begin{table}[!h]
\caption{Summary of the $\sigma_{\Delta T_{i,\text{ref}}}$ results when using cosmic-ray muons. The errors are statistical from the Gaussian fit.
}
\begin{center}
\begin{tabular}{|c||c||}
\hline
Bar ID		&$\sigma_{\Delta T_{i,\text{ref}}}$ (ns)			\\	
\hline
16	    & \num{0.40 \pm 0.005}	 \\
24		& \num{0.36 \pm 0.005} 	 \\
32		& \num{0.36 \pm 0.004} 	 \\ 
40		& \num{0.0  \pm 0.0}	     \\ 
48		& \num{0.38 \pm 0.004}	 \\ 
56		& \num{0.37 \pm 0.004}	 \\
64	    & \num{0.39 \pm 0.005}	 \\
\hline  
\end{tabular}
\label{tab:time_resolution}
\end{center}
\end{table}


\section{Discussion}
\label{sec:discussion}
A number of limitations of the current OS-SVSC prototype have motivated significant design changes for our next iteration. The presence of substantial electrical crosstalk described in Section~\ref{sec:crosstalk} has led to work toward custom arrays of individual SiPMs, as these allow more control over layout and component choices (e.g., connectors), which can help to limit electrical crosstalk effects. The ability to design increased spacing between pixels relative to the densely packed pixels of the commercial array also allows some opportunity for the mitigation of potential optical crosstalk. Such custom arrays can solve some other design issues as well. For example, by assembling SiPM pixels into individual $2\times8$ arrays, sets of 16 scintillator bars can be coupled as sub-modules. A full 64-bar detector could then be formed from pre-existing submodules. Having each bar externally accessible on a $2\times8$ submodule allows for improved optical coupling upon assembly, as each bar can be inspected and corrected if needed. Similarly, it is easy to access any bar on a submodule with sources, allowing all bars to be calibrated with the particle source techniques.

\begin{figure}[!h]
\centering
\begin{subfigure}{0.5\linewidth}
\centering
\includegraphics[width=\linewidth]{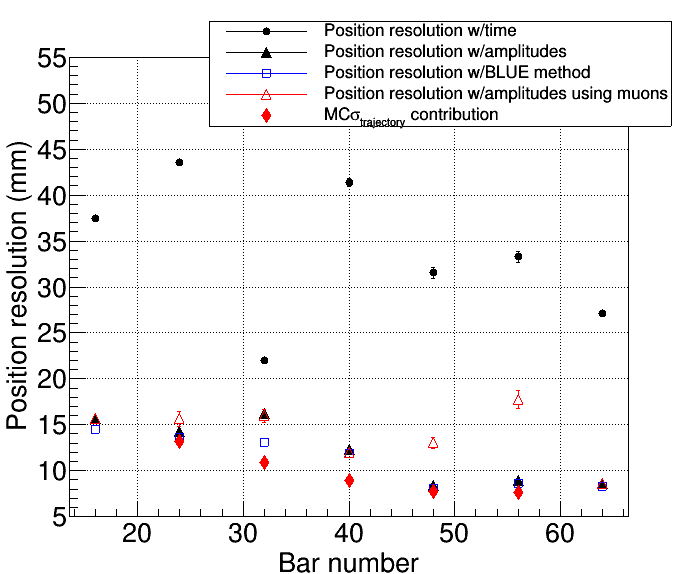}
\caption{}
\label{fig:resolutions}
\end{subfigure}
\begin{subfigure}{0.49\linewidth}
\centering
\includegraphics[width=\linewidth]{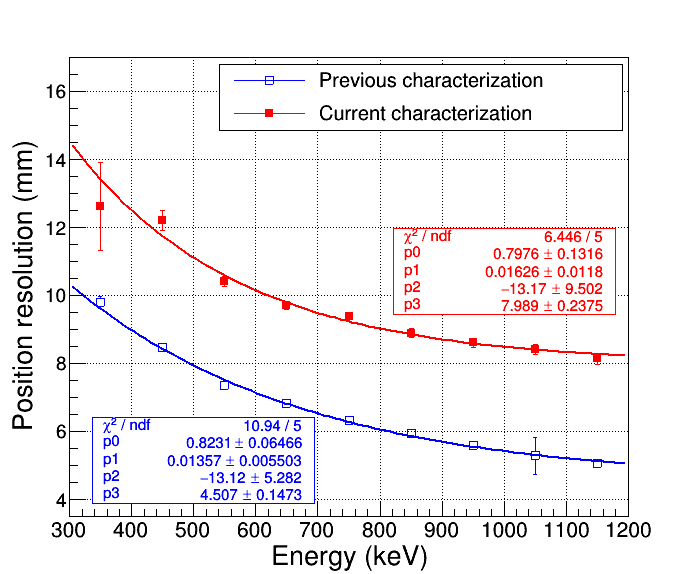}
\caption{}
\label{fig:energy_compare}
\end{subfigure}
\caption{(a) Position resolution results summarized for bars along one edge of the detector, including time-based, amplitude-based analyses, and the BLUE combination for source calibrations, and amplitude-based measurements for the muon calibration. $\sigma_{trajectory}$ is included as a contribution due to the error in the estimated muon trajectory, and (b) position resolution as a function of the energy of a bar calibrated in a previous effort reported on~\cite{bib:Sweany}, vs. bar 56 from the OS-SVSC prototype, both using the BLUE method. The threshold value (\SI{30}{ADC}) impacted the position resolution of bar 56 in the energy bin from \SIrange{300}{400}{keVee}, causing such an error.}
\end{figure}

Regarding the calibrations proposed for the current prototype, the muon-based energy ca\-li\-bra\-tion values from data were systematically lower than the simulation values by at least \SI{3}{\percent}. Nevertheless, the consistency among the energy calibration values lends confidence in the ability to calibrate the internal bars’ energy response without the need for source scans. The edge bars can still be calibrated using Compton edges of sources, and any disagreement between the methods can be taken as a systematic uncertainty on the energy calibrations.

The position reconstruction is limited to interactions that occurred within \SI{120}{mm} around each bar's center. A possible way to accurately reconstruct the interaction position near the ends of the bar would be to re-calibrate using a higher degree polynomial fit to the mean of the amplitude-based measurements with respect to the source position. For calibrating the position response, source scans remain an uncomplicated calibration method for bars along the detector's edge. Aggregate results for both timing- and amplitude-based resolutions, along with the BLUE method and the muon resolution results,  are shown in Fig.~\ref{fig:resolutions}. We found an increase in the position resolution of \SI{39.62}{\percent} when extrapolating the measurements using the BLUE method shown in this work with a power-law compared to the previous characterization effort, for depositions from \SIrange{300}{400}{keVee}~\cite{bib:Sweany}. Furthermore, an increase of \SI{54.66}{\percent} was found when comparing our measurements with extrapolated previous data for depositions from \SIrange{900}{1000}{keVee}, as is shown in Fig~\ref{fig:energy_compare}. This degradation may be attributed to several factors, including the use of different readout electronics, effects of electrical and optical crosstalk, and difficulty controlling the quality and uniformity of optical coupling while working with the full array of bars.  

The timing-based position resolutions turned out to be significantly worse than amplitude-based resolutions. The timing resolutions are significantly worse than those measured with the timing performance of the electronics, shown in Section \ref{sec:elec_calib}. Since timing calibrations were carried out on a single SCROD, we attribute much of this difference to jitter on the distribution of timing strobes from the CAJIPCI board that is used to synchronize both SCRODS. In addition, inconsistent digitization window-pairs between the SCRODs could also contribute.
The time resolutions from Section \ref{sec:time_resolution} showed time differences up to \SI{400}{ps}, a value that will limit the ability to analyze events with a very short time of flight between bars, in particular those where the neutron double-scatter occurs on immediately contiguous bars. Despite these limitations, we expect that we can still perform meaningful neutron double scatter imaging for events where the two interaction positions are at least \SI{6}{mm} apart, with the neutron traveling at the expected velocity of $\sim 0.05c$.  

\section{Conclusions}
We have described in detail the design of the OS-SVSC prototype. The chosen components (selected based on previous studies of scintillator bars, existing off-the-shelf solutions, and home-made pieces) allowed for the relatively fast delivery of a compact detector for testing.

Calibrations to the electronics allowed recording proper waveforms to calibrate energy and position using particle emission sources and energy, position, and time using a ``muon telescope''. The muon energy calibration method was repeatable and reliable despite a systematic decrease with respect to the particle emission source calibration. The position resolution calibration results show better performance when using amplitude-based than when using time-based measurements. Amplitude-based results are in a reasonable range for performing further work on neutron imaging.
Studies of muon-based position calibrations indicate a promising path toward full detector calibration to be used for interaction reconstructions. Namely, the edge bars can be calibrated for an amplitude-based response using sources. In turn, the calibrated edge bars are used to provide tracking information for muon trajectories to central bars to regenerate amplitude-based calibration curves. Results indicate position resolutions for such measurements to be less than \SI{17.74}{mm}. The comparison with the calibration using emission sources showed consistency. The time calibration method allowed for determining the time deviations between pairs of bars needed for double-neutron scatter detections. The method can be extrapolated further to the full detector by rotating it \ang{\sim 90} and setting the calibrated bars as the referenced ones.

Our collaboration is continuing to pursue measurements on this prototype OS-SVSC, intending to demonstrate neutron imaging.
A parallel effort is also underway to realize a second prototype with a modular design that addresses the limitations described in Section ~\ref{sec:discussion}.

\acknowledgments

Sandia National Laboratories is a multimission laboratory managed and operated by National Technology \& Engineering Solutions of Sandia, LLC, a wholly owned subsidiary of Honeywell International Inc., for the U.S. Department of Energy's National Nuclear Security Administration under contract DE-NA0003525. This paper describes objective technical results and analysis. Any subjective views or opinions that might be expressed in the paper do not necessarily represent the views of the U.S. Department of Energy or the United States Government. Document release number SAND2021-1287 O.

The authors would like to thank the US DOE National Nuclear Security Administration, Office of Defense Nuclear Nonproliferation Research and Development for funding this work. 

\end{document}